\documentclass[aps,superscriptaddress, nofootinbib, showpacs,10pt]{revtex4}
\usepackage{amsmath,amssymb,amsfonts, bm, natbib}
\usepackage{epsfig}
\usepackage{graphicx}
\usepackage{slashed}
\usepackage{color}
\usepackage{multirow}
\usepackage{caption}
\usepackage{subcaption}
\usepackage{slashed} 
\usepackage{mathrsfs}
\usepackage{soul}

\usepackage{calc}

\usepackage[font=small]{caption}

\def\beqn{\begin{eqnarray}}
\def\ba{\begin{array}{c}}
\def\bat{\begin{array}{cc}}
\def\bat{\begin{array}{cc}}
\def\ea{\end{array}}
\def\bat{\begin{array}{cc}}
\def\batt{\begin{array}{ccc}}
\def\eeqn{\end{eqnarray}}

\def\ta{{\tilde{\alpha}}}
\def\Hpm{{H^{\pm}}}

\newcommand{\be}{\begin{equation}}
\newcommand{\ee}{\end{equation}}
\newcommand{\bel}[1]{\be\label{#1}}
\newcommand{\fr}{\dfrac}
\newcommand{\lt}{\left}
\newcommand{\rt}{\right}

\makeatletter
\newcommand*{\rom}[1]{\expandafter\@slowromancap\romannumeral #1@}
\makeatother

\definecolor{gray}{rgb}{0.4,0.4,0.4}
\newcommand{\cR}{\mathcal{R}}

\begin{document}
\title{Loop induced $H^\pm \rightarrow  W^\pm Z$ decays in the aligned two-Higgs-doublet model}
\author{Gauhar Abbas}
\email{gauhar.phy@iitbhu.ac.in}
\affiliation{Department of Physics, Indian Institute of Technology (BHU), Varanasi 221005, India}

\author{Diganta Das}
\email{diganta99@gmail.com}
\affiliation{Department of Physics and Astrophysics, University of Delhi, Delhi 110007, India}

\author{Monalisa Patra}
\email{monalisa.patra@ijs.si}
\affiliation{Jo\v{z}ef Stefan Institute, Jamova 39, P. O. Box 3000, 1001 Ljubljana, Slovenia}

\begin{abstract}
We present a complete one-loop computation of the $H^\pm \rightarrow  W^\pm Z$  decay in the aligned two-Higgs-doublet model.  The constraints from the electroweak precision observables, perturbative unitarity, vacuum stability and flavour physics are all taken into account along with the latest Large Hadron Collider searches for the charged Higgs.  It is observed that a large enhancement of the branching ratio can be obtained in the limit where there is a large splitting between the charged and pseudo-scalar Higgs masses as well as for the largest allowed values of the alignment parameter $\varsigma_u$.  We find that the maximum possible branching ratio in the case of a large mass splitting between $m_\Hpm$ and $m_A$ is $\approx 10^{-3}$ for $m_\Hpm \in (200,700)$ GeV which is in the reach of the high luminosity phase of the Large Hadron Collider. 
\end{abstract}

%\pacs{14.60.Pq, 11.10.Hi, 11.30.Hv, 12.15.Lk}

\maketitle

\section{Introduction}
Following the discovery of a Higgs-like particle at the Large Hadron Collider (LHC), we are a step closer to understanding the electroweak symmetry breaking (EWSB) mechanism in the Standard Model (SM). This discovery however raises one important question, that is, whether the Higgs-like particle is indeed the Higgs of the SM or a component of an extended scalar sector corresponding to a richer EWSB scenario than in the SM.   
One of the simplest beyond SM scenarios is  the two-Higgs-doublet model (2HDM) where the SM Higgs doublet is supplemented with one additional scalar doublet~\cite{Gunion:1989we,Branco:2011iw}. There are many motivations to introduce extra Higgs doublets, for example to explain the  electroweak baryogenesis~\cite{Turok:1990zg}, top-bottom mass hierarchy~\cite{Hashimoto:2004xp}, and neutrino mass generation~\cite{Zee:1980ai}, to name a few.  The discerning feature of the extension with one extra Higgs doublets is that it leads to four additional scalar particles beyond the SM, namely, two charged scalars and two neutral scalars. Various properties of these additional scalars can be probed through precise determinations of the Higgs properties such as its mass, production cross section, and its decays involving the SM-like Higgs~\cite{Heinemeyer:2005gs}.  The direct searches of these scalar particles at the LHC could help us in acquiring an understanding  of the scalar sector of a more fundamental underlying theory.

The charged Higgs ($H^\pm$) is one of the new particles of the extended Higgs sector of the 2HDM, and if such particle exists, its direct detection could lead us to a better understanding of the extended scalar sector. The charged Higgs is currently been searched at the LHC through different production and decay modes~\cite{Akeroyd:2016ymd}. The $tb$ decay mode is considered in the search of a heavy charged Higgs whereas the preferred decay mode channel for light charged Higgs searches is the $\tau\nu_\tau$ channel.  

Among the various decay channels of the charged Higgs, the $W^\mp Z$  decay mode is quite interesting because the $H^\pm W^\mp Z$ vertex does not occur at tree level in general multi-Higgs doublet models, in contrast to more exotic scalar sectors (e.g., triplets) where this decay can occur at tree level~\cite{Godbole:1994np}. The absence of this tree-level vertex is due to the weak isospin symmetry of the scalar kinetic terms~\cite{Grifols:1980uq,Rizzo:1989ym}. The $H^\pm W^\pm Z$ vertex in  2HDM is therefore loop-induced,  however it is well known that an observable enhancement in the magnitude of the vertex can come from non-decoupling effects of particles running in the loop. These are in particular the interactions which break custodial symmetry, for instance, the top and bottom quark-loop contributions to the $H^\pm W^\pm Z$  vertex  show a quadratic dependence on the top quark mass \cite{CapdequiPeyranere:1990qk}\footnote{The multi-Higgs doublet model being $U(1)_{em}$ symmetric, the vertex $H^\pm W^\mp \gamma$ is also loop induced and receives only logarithmic mass effects. Therefore the $H^\pm \rightarrow W^\pm \gamma$ amplitude is not sensitive to the non-decoupling effects.}. In the context of 2HDM of type \rom{2},  it is shown in Refs.~\cite{Kanemura:1997ej,Kanemura:1999tg} that an enhancement of $H^\pm \rightarrow  W^\pm Z$ is possible due to the non-decoupling effect of the heavy Higgs bosons, \emph{i.e.,} a large mass difference between the CP-odd neutral scalars and the charged Higgs that breaks the custodial symmetry.  Thus, the $H^\pm W^\pm Z$ vertex  has nontrivial  consequences in the context of custodial symmetry. This decay channel has also been studied in the context of three Higgs doublet models \cite{Moretti:2015tva}. 

In the most general version of 2HDMs there are large flavour changing neutral current (FCNC) interactions which are in conflict with various flavour data. This problem is usually avoided by the natural flavour conservation (NFC) hypothesis, implementing a discrete $Z_2$ symmetry that allows only one scalar field to couple to a given type of right-handed fermion~\cite{Glashow:1976nt,Paschos:1976ay} and hence evades tree-level FCNC. In the aligned two-Higgs-doublet model (A2HDM) the FCNC problem is solved in a more general way by aligning the Yukawa matrices in the flavour space \cite{Pich:2009sp}. It is based on the assumption that the Yukawa matrices coupled to a given right-handed fermion have the same flavour structure. These matrices can then be diagonalized simultaneously leading to no FCNCs at tree level.  The scalar sector in the A2HDM is similar to the scalar sector of the most general 2HDMs whereas the Yukawa sector is parametrized in terms of three complex couplings $\varsigma_{u,d,\ell}$, known as the alignment parameters. The A2HDM can be considered as a relatively general framework, from which all the known versions of the 2HDMs can be recovered under different limits of the alignment parameters. Phenomenological analyses of the A2HDM taking into account the latest LHC results and flavour physics observables can be found in Refs.~\cite{Jung:2010ab,Jung:2010ik,Jung:2012vu,Celis:2012dk,Jung:2013hka,Li:2014fea,Dekens:2014jka,Altmannshofer:2012ar,Bai:2012ex,Celis:2013rcs,Barger:2013ofa,Lopez-Val:2013yba,Duarte:2013zfa,Celis:2013ixa,Wang:2013sha,Abbas:2015cua,Enomoto:2015wbn,Hu:2016gpe,Gori:2017qwg,Penuelas:2017ikk}.

In this work we study the $H^\pm \rightarrow  W^\pm Z$ decay within the framework of the CP-conserving A2HDM. We take into account the most recent limits from the LHC along with theoretical constraints such as vacuum stability, perturbative unitarity and experimental bounds from charged Higgs searches at LEP and flavour physics. In the case of the 2HDM of type \rom{2},  it was shown that with the soft-breaking parameter being small,  a large mass difference between the charged and the CP-odd scalars in the non-decoupling limit leads to an enhanced contribution to the decay width from the scalar loop diagrams~\cite{Kanemura:1997ej,Kanemura:1999tg}. However, in the A2HDM the scalar loop diagrams are proportional to the quartic couplings which are either independent parameters or are functions of masses.  Therefore a large values of these independent quartic couplings along with a large mass splitting in the Higgs sector leads to an enhanced contribution from the Higgs-boson loop diagrams in our case. 

The paper is organized as follows. We briefly describe the A2HDM in section \ref{sec2}. The theoretical and  experimental constraints on the parameter space of the A2HDM are discussed in Sec.~\ref{sec3} . We evaluate the decay $H^\pm \rightarrow  W^\pm Z$  and the relevant branching ratios in the A2HDM in section \ref{sec4} and in section \ref{sec5} we present  the  results of  the  LHC production cross section for the processes $gb \rightarrow H^+\bar{t}$ and single charged Higgs production through $WZ$ fusion with the subsequent decay of $H^\pm$ to  $W^\pm Z$.  We finally summarize our results in section \ref{sec6}. The analytical results of the various diagrams contributing to the decay amplitude are listed in the appendices.

\section{The aligned two-Higgs doublet model}
\label{sec2}
The two complex scalar doublets of the A2HDM in the Higgs basis, where only one doublet acquires a vacuum expectation value (VEV) can be written as~\cite{Pich:2009sp}
\begin{equation}  \label{Higgsbasisintro}
\Phi_1=\left[ \begin{array}{c} G^+ \\ \frac{1}{\sqrt{2}}\, (v+S_1+iG^0) \end{array} \right] \; ,
\qquad\qquad
\Phi_2 = \left[ \begin{array}{c} H^+ \\ \frac{1}{\sqrt{2}}\, (S_2+iS_3)   \end{array}\right] \; ,
\end{equation}
where  $v = ( \sqrt{2} G_F )^{-1/2} \simeq 246$~GeV, $G^{0, \pm}$ denote the would-be Goldstone bosons, and $H^{\pm}$ are the charged Higgs.  The three neutral Higgs bosons  are denoted by $\varphi_j^0(x)=\{h(x),H(x),A(x)\}$ and they are related to the $S_i$ fields by the transformation  $\varphi_j^0 = \cR_{jk}  S_k $.  The $\cR$ matrix is  orthogonal and diagonalizes the mass terms in the scalar potential~\cite{Celis:2013rcs}. 
The most general scalar potential of the 2HDM is of the form
%%%%%%%%%%%%%%%%%
\begin{eqnarray}\label{eq:scalar_potential}
V & = &  \mu_1\; \Phi_1^\dagger\Phi_1\, +\, \mu_2\; \Phi_2^\dagger\Phi_2 \, +\, \left[\mu_3\; \Phi_1^\dagger\Phi_2 \, +\, \mu_3^*\; \Phi_2^\dagger\Phi_1\right]
\nonumber    \\[0.2cm] & + & \lambda_1\, \left(\Phi_1^\dagger\Phi_1\right)^2 \, +\, \lambda_2\, \left(\Phi_2^\dagger\Phi_2\right)^2 \, +\,
\lambda_3\, \left(\Phi_1^\dagger\Phi_1\right) \left(\Phi_2^\dagger\Phi_2\right) \, +\, \lambda_4\, \left(\Phi_1^\dagger\Phi_2\right) \left(\Phi_2^\dagger\Phi_1\right)
\nonumber \\[0.2cm] & + & \biggl[  \left(\lambda_5\; \Phi_1^\dagger\Phi_2 \, +\,\lambda_6\; \Phi_1^\dagger\Phi_1 \, +\,\lambda_7\; \Phi_2^\dagger\Phi_2\right) \left(\Phi_1^\dagger\Phi_2\right)
\, +\, \mathrm{h.c.}\,\biggr]\,.
\end{eqnarray}
Due to Hermiticity, all parameters appearing in $V$ are real except $\mu_3, \lambda_5, \lambda_6$ and $\lambda_7$ that introduce additional source of CP violation. To reduce the number of independent parameters in our analysis we limit ourselves to the CP conserving case, so that $\mu_3, \lambda_5, \lambda_6$ and $\lambda_7$ are real. 
The  minimization of the scalar potential leads to the following relations:  
\begin{equation}
\mu_1\; =\; -\lambda_1\, v^2\, ,
\qquad\qquad\qquad
\mu_3\; =\; -\frac{1}{2}\,\lambda_6\, v^2\, ,
\end{equation}
%%%%%%%%%%%%%%%%
with the charged Higgs mass being  given by
%%%%%%%%%%%%%%%%%%%%%%%
\begin{equation}\label{eq:mHpmMu2Lam}
m_{H^\pm}^2\; =\; \mu_2 + \frac{1}{2}\,\lambda_3\, v^2 \,.
\end{equation}
%%%%%
In the CP-conserving limit,  the CP-odd field $A$ directly corresponds  to $S_3$ and the physical neutral Higgs  bosons are related to $S_1$ and $S_2$ through the following transformation:
\begin{equation}\label{eq:rot_matrix}
\left(\ba h\\ H\ea\right)\; = \;
\left[\bat \cos{\tilde\alpha} & \sin{\tilde\alpha} \\ -\sin{\tilde\alpha} & \cos{\tilde\alpha}\ea\right]\;
\left(\ba S_1\\ S_2\ea\right) \, .
\end{equation}
When $m_h \leqslant m_H$, the angle $\tilde \alpha$ is given by the following relations
\begin{eqnarray}
\sin 2 \tilde \alpha\; =\;   \frac{-2 \lambda_6 v^2  }{  m_H^2 - m_h^2   }    \,,
\qquad\qquad
\cos 2 \tilde \alpha\; =\;   \frac{ m_A^2 + 2 (\lambda_5 - \lambda_1) v^2   }{  m_H^2 - m_h^2   }  \,.
\end{eqnarray}
The range of the mixing angle $\tilde \alpha$ is constrained to $0 \leqslant \tilde \alpha < \pi$  through a phase redefinition of the CP-even fields.
The scalar  masses  in the CP-conserving limit are given as
%%%%
\begin{equation}
m_h^2\; =\;\frac{1}{2}\,\left( \Sigma-\Delta\right) ,
\quad\quad
m_H^2\; =\;\frac{1}{2}\,\left( \Sigma+\Delta\right) ,
\quad\quad
m_A^2 \; =\; m_{H^\pm}^2\, +\, v^2\,\left(\frac{\lambda_4}{2} - \lambda_5\right) ,
\end{equation}
%%%%
with
%%%%
\begin{eqnarray}
\Sigma & \;=\;&    m_{H^{\pm}}^{2}   +  \left( 2 \lambda_1 +  \frac{ \lambda_4 }{2}  + \lambda_5 \right)  v^2\, ,  \quad    \Delta  = \sqrt{\left[  m_A^2 + 2  (\lambda_5- \lambda_1) v^2    \right]^2 + 4 v^4 \lambda_6^2}\, .
%\\[0.2cm] \label{eq:Delta}
%\Delta & \;=\;&\sqrt{\left[  m_A^2 + 2  (\lambda_5- \lambda_1) v^2    \right]^2 + 4 v^4 \lambda_6^2}\, .
\end{eqnarray}
%%%%
The Yukawa Lagrangian in the A2HDM in terms of the fermion mass-eigenstates is written as~\cite{Pich:2009sp}
%%%
\beqn \label{lagrangianY}
 \mathcal L_Y & = &  - \frac{\sqrt{2}}{v}\; H^+ \Bigl\{ \bar{u} \left[ \mathrm{\varsigma}_d\, V_{\rm CKM} m_d \, P_R - \mathrm{\varsigma}_u\, m_u^\dagger V_{\rm CKM}  \, P_L \right]  d\, + \, \mathrm{\varsigma}_\ell\, \bar{\nu}~m_\ell \, P_R \ell \Bigr\}
\nonumber \\[0.2cm]
& & -\,\frac{1}{v}\; \sum_{\mathrm{\varphi}^0_i, f}\, y^{\varphi^0_i}_f\, \varphi^0_i  \; \left[\bar{f}\,  m_f \, P_R  f\right]
\; + \;\mathrm{h.c.} \, ,
\eeqn
%%%%
where $P_{L,R} = (1 \mp \gamma_5)/2$ are the chirality projection operators, $m_{f=u,d,\ell}$ are the fermion masses, and  $V_{\rm CKM}$ is the Cabibbo--Kobayashi--Maskawa (CKM) matrix element.  The neutral Higgs couplings are given by
\begin{equation}  \label{yukascal}
y_{d,\ell}^{\mathrm{\varphi}^0_i} = \cR_{i1} + (\cR_{i2} + i\,\cR_{i3})\,\mathrm{\varsigma}_{d,\ell}  \, ,
\qquad\qquad
y_u^{\mathrm{\varphi}^0_i} = \cR_{i1} + (\cR_{i2} -i\,\cR_{i3}) \,\mathrm{\varsigma}_{u}^* \, .
\end{equation}
The  parameters $\mathrm{\varsigma}_{f}$ ($f=u,d,\ell$) represent alignment conditions in the flavour space and are family-universal complex quantities leading to new sources of CP violation beyond the CKM matrix. We consider these parameters to be real for our analysis. All the known versions of the  2HDM with natural flavour conservation can be recovered by taking particular limits of the aligned  parameters as shown in table~\ref{tab:NFC}.  The most stringent constraints on the modulus of the aligned parameters come from flavour physics to be discussed in the next sections.  
%%%%%%%%%%%%%%%%%%%%%%%%%%%%%%%%%%%%%%%%%%%%%%%%%%%%%%%%%%%%%%%%%%%%%%%%%%
\begin{table}[htb]
\begin{center}
\vspace{0.2cm} \tabcolsep 0.15in
\begin{tabular}{|c|c|c|c|}
\hline 
Model & $\varsigma_d$ & $\varsigma_u$ & $\varsigma_\ell$  \\[0.1cm]
\hline
Type I  & $\cot{\beta}$ &$\cot{\beta}$ & $\cot{\beta}$ \\[0.1cm]   
Type II & $-\tan{\beta}$ & $\cot{\beta}$ & $-\tan{\beta}$ \\[0.1cm]
Type X (lepton specific)  & $\cot{\beta}$ & $\cot{\beta}$ & $-\tan{\beta}$ \\[0.1cm] 
Type Y (flipped) & $-\tan{\beta}$ & $\cot{\beta}$ & $\cot{\beta}$ \\[0.1cm]
\hline
\end{tabular}
\caption{\it \small The couplings $\varsigma_f$ in various types of two-Higgs-doublet models with $Z_2$ symmetry.}
\label{tab:NFC}
\end{center}
\end{table}
%%%%%%%%%%%%%%%%%%%%%%%%%%%%%%%%%%%%%%%%%%%%%%%%%%%%%%%%%%%%%%%%%%%%%%%%%%

%
%%%%%%%%%%%%%%%%%%%%%%%%%%%%%%%%%%%%%%%%%%%%%%%%%%%%%%%%%%%%%%%%%%%%%%%%%%%%%%%%%%%%%%%%%%%%%
%%%%%%%%%%%%%%%%%%%%%%%%%%%%%%%%%%%%%%%%%%%%%%%%%%%%%%%%%%%%%%%%%%%%%%%%%%%%%%%%%%%%%%%%%%%%%
\section{Theoretical and experimental constraints}\label{sec3}
In this section we explore the various theoretical and experimental constraints on the parameter space of the CP-conserving A2HDM. In this limit there are 11 real free parameters which include $\mu_2$, the couplings $\lambda_i(i=1...7)$, and the three alignment parameters $\mathrm{\varsigma}_{u,d,\ell}$.
% 
%and the counter-term $C(\mu)$\footnote{The higher-order corrections de-stabilize the alignment condition but the flavour structure of the A2HDM  strongly constrain the possible FCNC effects. The one loop FCNC effect arising due to the quantum corrections is given by the following equation~\cite{Jung:2010ik}:
%\begin{align} \label{fcnc:ct}
%\mathcal{L}_{\mbox{\scriptsize{FCNC}}} \;=\;  & \frac{C}{  4 \pi^2 v^3 }\,   (1 + \varsigma_u^* \varsigma_d )     \sum_j  \, \varphi_j^0 \, \biggl\{    (  \mathcal{R}_{j2}  + i \mathcal{R}_{j3} ) (  \varsigma_d - \varsigma_u ) \Bigl[ \bar d_L \, V^{\dag}  M_u M_{u}^{\dag} V M_d \, d_R    \Bigr]   \nonumber \\
%& - (\mathcal{R}_{j2} - i \mathcal{R}_{j3})  ( \varsigma_d^* - \varsigma_u^* ) \Bigl[  \bar u_L \, V M_d M_d^{\dag}   V^{\dag}  M_u u_R  \Bigr]  \biggr\} + \mathrm{h.c.} \,,
%\end{align}
%where the renormalized coupling constant $C$ can be written as  $C(\mu) = C_{R}(\mu)  + \frac{1}{2}  \left\{ \frac{  2 \mu^{D-4}}{D-4}   + \gamma_E - \ln(4\pi) \right\}$, with  $C_R(\mu) = C_R(\mu_0)  - \ln(\mu/\mu_0)$~\cite{Li:2014fea}.  Here $\gamma_E\simeq 0.577$ is the Euler constant, and $\mu$ is an arbitrary renormalization mass scale. If we assume that alignment is exact at a high energy scale $\Lambda_A$ so that $C_R(\Lambda_A)=0$, then we obtain $C_R(\mu) = \ln(\Lambda_A/\mu)$.  Since $C(\mu)$ does not contribute to the process studied here, we work with 11 parameters.}. 
%
Four of the parameters of the scalar potential can be expressed in terms of the physical scalar masses and the mixing angle $\tilde{\alpha}$  and are given by 
\begin{eqnarray}\label{eq:parameters}
	\lambda_1 &=& \frac{1}{2v^2}(m_h^2 \cos^2\ta + m_H^2 \sin^2\ta )\, ,\quad \lambda_4 = \frac{1}{v^2}( m_h^2 \sin^2\ta + m_H^2 \cos^2\ta + m_A^2 - 2m^2_\Hpm )\, ,~~~\\
	\lambda_5 &=& \frac{1}{2v^2}( m_h^2 \sin^2\ta + m_H^2 \cos^2\ta - m_A^2 )\, ,\quad \lambda_6 = -\frac{1}{v^2} (m_H^2 - m_h^2) \cos\ta \sin\ta\, .
\end{eqnarray}
Taking into account the above relations along with Eq.~(\ref{eq:mHpmMu2Lam}) leads us to work with a set of parameters that can be related to the physical masses $m_h, m_A, m_H, m_\Hpm$, the mixing parameter $\cos\ta$, three couplings $\lambda_{2,3,7}$, and the Yukawa parameters $\mathrm{\varsigma}_{u,d,\ell}$.  We have fixed $m_h = 125.5$ GeV in our calculation with the assumption that  the scalar boson observed by the ATLAS \cite{Aad:2012tfa} and the CMS collaborations \cite{Chatrchyan:2012xdj} correspond to the lightest CP-even state $h$ in the A2HDM. We also set $\cos\tilde{\alpha} = 0.95$ in order to ensure that the couplings of $h$ to the gauge bosons, $\lambda^h_{WW}$ and $\lambda^h_{ZZ}$, remain consistent with the LHC data.

\begin{figure}[htb]
\centering
 \includegraphics[width=11cm, height=7cm]{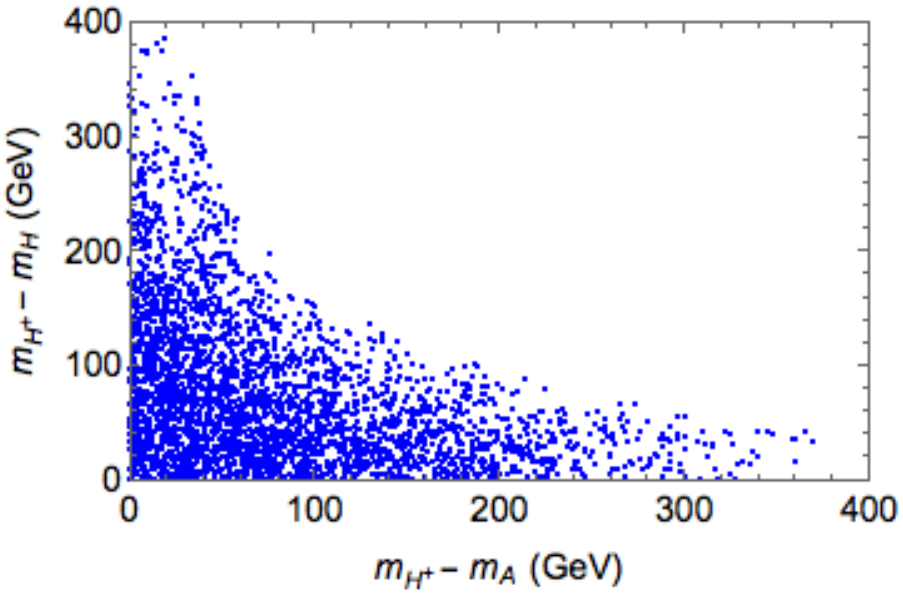}
 \caption{The region of scalar mass splitting in $|m_{H^\pm}-m_H|$ vs $|m_{H^\pm}-m_A|$ plane, allowed by the Higgs signal strength in the di-photon channel, perturbativity bounds on the quartic scalar couplings $|\lambda_{2,3,7}|<4\pi$, stability of scalar potential, unitarity of $S$-wave scattering amplitudes, and the electroweak precision data.}
 \label{fig:mass_diff}
\end{figure}

The loop induced $h\to\gamma\gamma$ decay width receives a contribution from the charged Higgs, making this process sensitive to $\lambda_{3,7}, m_{H^\pm}$ and  $\tilde{\alpha}$. The Higgs signal strength in the diphoton channel has been measured at the LHC, with the latest results from ATLAS \cite{Khachatryan:2016vau} and CMS \cite{Khachatryan:2014jba} being $\mu^h_{\gamma\gamma}=1.17^{+0.28}_{-0.26}$ and $\mu^h_{\gamma\gamma}=1.12\pm 0.24$ respectively. The Higgs production cross section being the same as in the SM, the signal strength in the A2HDM reads \cite{Celis:2013rcs,Celis:2013ixa}
\begin{eqnarray}
	\mu^h_{\gamma\gamma} &=& \frac{\sigma(pp\to h)\times \text{Br}(h\to 2\gamma)}{\sigma(pp\to h)_{\rm SM}\times \text{Br}(h\to 2\gamma)_{\rm SM}} \simeq (1 - 0.15 ~C^h_{H^\pm})^2,\quad \mathrm{with} \\
	\label{eq:ChH}
	C^h_{H^\pm} &=& \frac{v^2}{2m^2_\Hpm} x_\Hpm \lambda^h_{H^+H^-} \Bigg( -1 +  x_\Hpm \arcsin^2(\frac{1}{\sqrt{x_\Hpm}})  \Bigg)\, ,
\end{eqnarray}
where $x_\Hpm = 4m_\Hpm^2/ m_h^2$.  We have imposed the condition that $\mu^h_{\gamma\gamma}$ in our case should lie within the 2$\sigma$ range of the experimental measurements. Additionally, the $\lambda^h_{H^+H^-}$ coupling ($\simeq \lambda_3 \cos\tilde{\alpha} + \lambda_7 \sin\tilde{\alpha}$) in Eq.~(\ref{eq:ChH}) is required to be less than 4$\pi$ to make sure the validity of perturbation theory. However for a light charged Higgs, this cubic coupling receives a sizable one loop scalar contribution \cite{Celis:2013rcs} 
\begin{eqnarray}
(\lambda^h_{H^+H^-})_{\mathrm{ eff}} &=& \lambda^h_{H^+H^-} (1 + \Delta), ~ \mathrm{with}~\Delta=\frac{v^2 (\lambda^h_{H^+H^-} )^2}{16 \pi^2 m_{H^\pm}^2}\mathcal{Z}\left(\frac{m_h^2}{m_{H^\pm}^2}\right),
\end{eqnarray}
with
\begin{eqnarray}
\mathcal{Z}(X)&=& \int_0^1dy~\int_0^{1-y}dz \left[(y+z)^2+X(1-y-z-yz)\right]^{-1}.
\end{eqnarray}
Since a large correction to $\lambda^h_{H^+H^-}$ could invalidate the perturbation theory, at most $50\%$ corrections are allowed, \emph{i.e.,} $\Delta\leq 0.5$. The other theoretical bounds considered are the perturbativity bounds on the quartic scalar couplings $|\lambda_{2,3,7}| < 4\pi$, the requirement of the stability of the scalar potential~\cite{Branco:2011iw}, and the unitarity of the $S$-wave scattering amplitudes of the scalars~\cite{Ginzburg:2005dt}. Additionally, the electroweak precision tests provide important constraints on the parameters of the A2HDM. The mass splittings between the additional scalars of the A2HDM are constrained by the S, T U parameters~\cite{Peskin:1991sw}. It was shown in Ref.~\cite{Celis:2013ixa} that in order to satisfy the precision electroweak constraints the mass differences $|m_{H^\pm}-m_H|$ and $|m_{H^\pm}-m_A|$ cannot be both larger than $v$ at the same time.  Taking into account all these constraints, we perform a scan in the $m_A, m_H, m_\Hpm, \lambda_{2,3,7}$ parameter space. The points for the scan are generated in the intervals
\begin{eqnarray}
&& \lambda_2 \in (0, 4\pi),\qquad  \lambda_3 \in (-4\pi, 4\pi),\qquad \lambda_7 \in (-4\pi, 4\pi),\qquad m_\Hpm \in (180~\mathrm{GeV}, 1~\mathrm{TeV}),\nonumber \\
&& m_{H}\in (180~\mathrm{GeV}, 1~\mathrm{TeV}),\qquad m_{A}\in \left(180~\mathrm{GeV}, 1~\mathrm{TeV}\right).
\end{eqnarray}
The allowed parameter space for the scalar mass differences $m_{H^\pm}-m_A$ and $m_{H^\pm}-m_H$ is shown in Fig.~\ref{fig:mass_diff}. This shows that there can be three possible scenarios
%In order to analyze the effect of the mass differences on the $H^\pm \rightarrow W^\pm Z$ decay width, we will present our results for following scenarios
%
\begin{itemize}
\item Case 1 : $|m_{H^\pm}-m_A| \geq$ 200 GeV and  $|m_{H^\pm}-m_H| \leq$ 40 GeV,
\item Case 2 :  $|m_{H^\pm}-m_H| \geq$ 200 GeV  and $|m_{H^\pm}-m_A| \leq$ 40 GeV,
\item Case 3 : $|m_{H^\pm}-m_H|$ $\simeq$  $|m_{H^\pm}-m_A|$ $\leq$ 40 GeV
\end{itemize}
We will later discuss the decay width of $H^\pm \rightarrow W^\pm Z$ in the context of these three scenarios.  After discussing the constraints on the couplings and the physical masses, we will now study the constraints on the alignment parameters in the next sections.

\subsection{Impact on $\varsigma_{u,d,\ell}$ from flavour observables and direct LHC searches of charged Higgs}\label{sec3A}

 Firstly we discuss the constraints currently available on the alignment parameters from flavour physics. The inclusive $B\to X_{s,d}\gamma$ branching ratio constrains the  $\mathrm{\varsigma}_{u}-\mathrm{\varsigma}_{d}$ parameter space. The alignment parameter for the up-quark is additionally constrained from the $B_{s,d}^{0}-\bar{B}^{0}_{s,d}$ mixing and from the $Z\rightarrow b\bar{b}$ decay width.  The   $Z\rightarrow b\bar{b}$ branching ratio leads to a linear dependence on the charged Higgs mass which implies~\cite{Jung:2010ik}
\begin{equation}
|\varsigma_u| < 0.72 + 0.0024~m_\Hpm~\mathrm{GeV}^{-1}\, ,~~~{\rm (at~95\%~CL)}.
\end{equation}
The other two alignment parameters $\varsigma_{d,\ell}$ are constrained with the requirement that the Yukawa couplings should remain within the perturbative regime, ($\sqrt{2}\varsigma_{d,\ell}m_{d,\ell}/v<1)$, leading to absolute upper bounds $|\varsigma_d|<50$ and  $|\varsigma_\ell|<100$. For our analysis, we vary the alignment parameters  in the following region taking into account the above constraints as well as the flavour constraints from radiative inclusive $B\to X_{s,d}\gamma$  decays~\cite{Jung:2010ik},
\begin{eqnarray}\label{eq:align_range}
&& \varsigma_u \in (-3, 3),\qquad  \varsigma_d \in (-50, 50),\qquad \varsigma_\ell \in (-100,100)\, .
\end{eqnarray}

Apart from the bounds considered before, the direct searches of new scalars at the LHC and LEP provide additional constraints on the model parameters. Here we will consider the constraints coming from the charged Higgs searches. The LEP collaborations searched for a charged Higgs in the $e^+e^- \rightarrow H^+H^-$ channel with the charged Higgses reconstructed from $H^+ \rightarrow c\bar{s}$ and $\tau^+\nu_\tau$. The non-observation of any signal at LEP collaboration puts a lower bound on the charged Higgs mass: $m_\Hpm \geq 78.6$ GeV~\cite{Searches:2001ac} at 95\% CL in the 2HDM of type  \rom{2}. 

The LHC has searched for a light charged Higgs in the $t\rightarrow H^\pm b$ channel and has excluded $m_\Hpm \in$ (80,160) GeV~\cite{CMS:2016szv}. The LHC collaboration has also looked for a heavy charged Higgs in the $pp \rightarrow t(b)H^\pm$ process with $H^\pm \rightarrow \tau^\pm \nu_\tau$~\cite{CMS:2016szv,ATLAS:2016grc}, $H^\pm \rightarrow tb$~\cite{ATLAS:2016qiq}, $H^\pm \rightarrow W^\mp Z$~\cite{Sirunyan:2017sbn} and has given a model independent limit on $\sigma(pp \rightarrow t(b)H^\pm) \times BR(H^\pm \rightarrow \tau^\pm \nu_\tau,tb)$ as a function of $m_H^\pm$. We use this limit to constrain the alignment parameters, in addition to the constraints from flavour observables discussed above.

To implement the LHC bounds we calculate the process $\sigma(pp \rightarrow t(b)H^\pm) \times BR(H^\pm \rightarrow \tau^\pm \nu_\tau,tb)$ in the A2HDM in \texttt{Madgraph}~\cite{Alwall:2011uj}. The dependence of the alignment parameters on the production cross section $\sigma(pp \rightarrow t(b)H^\pm)$ comes through the vertex $\lambda^{H^\pm}_{tb} = \varsigma_u m_t P_L - \varsigma_d m_b P_R$. The decay widths of the charged Higgs to $tb$ and $\tau\nu_\tau$ are also sensitive to the alignment parameter and are given by

\begin{eqnarray}
\Gamma(H^\pm\rightarrow  tb) &=& \frac{3\lambda^{1/2}(m_\Hpm^2, m_t^2,m_b^2)}{8\pi  v^2 m_\Hpm^3}\left[(m_\Hpm^2-m_t^2)\left(m_b^2 \varsigma_d^2+m_t^2\varsigma_u^2\right) \right.\nonumber \\
&& \left. 
-m_b^2\left(m_b^2 \varsigma_d^2+m_t^2\varsigma_u^2\right)+4 m_b^2m_t^2 \varsigma_d \varsigma_u \right] \nonumber \\
\Gamma(H^\pm \rightarrow \tau^\pm \nu_\tau) &=& \frac{m_\tau^2 \varsigma_\ell^2}{8\pi  v^2 m_\Hpm^3} \left[m_\Hpm^2-m_\tau^2\right]^2,
\end{eqnarray}
with $\lambda (a,b,c) = (a - b - c)^2 - 4 b c$. 

In our numerical analysis we have also added the branching ratios into $cs, ud$ and $\mu \nu_\mu$ which can be trivially obtained from the above formulas. In the limit  $m_\Hpm > m_W+ m_{\varphi_i}$ where $\varphi_i = h, H, A$, the decay $H^\pm \rightarrow W^\pm \varphi_i$ is kinematically allowed. The corresponding decay width is given by

\begin{eqnarray}
\Gamma(H^\pm\rightarrow  W^\pm \varphi_i) &=& \frac{\lambda^{3/2}(m_\Hpm^2, m_W^2,m_{\varphi_i}^2)}{16\pi v^2 m_\Hpm^3} R_{i}^2,
\end{eqnarray} 
with $R_{h(H)}=\sin(\cos)\tilde{\alpha}$ and $R_A$ = 1.

Compared to the other fermionic modes, $H^\pm\to tb$ has the dominant branching ratio for sizable alignment parameters since it depends on the mass of the top quark.  Therefore, the $tb$ decay channel at the LHC can be used to further constrain the $\varsigma_u - \varsigma_d$ parameter space.  The decay width $\Gamma(H^\pm\rightarrow  W^\pm \varphi_i)$ being independent of the alignment parameters, when kinematically allowed the BR$(H^\pm\rightarrow  W^\pm \varphi_i)$  will be dominant in the limit of small values of alignment parameters. Note that, since $|\varsigma_\ell|<100$ \cite{Penuelas:2017ikk,Jung:2012vu} is a very weak constraint, the BR$(H^\pm \rightarrow \tau^\pm \nu_\tau)$ could dominate for a light charged Higgs.  
\begin{figure}[htb]
	\centering
	\includegraphics[width=7.5cm, height=6cm]{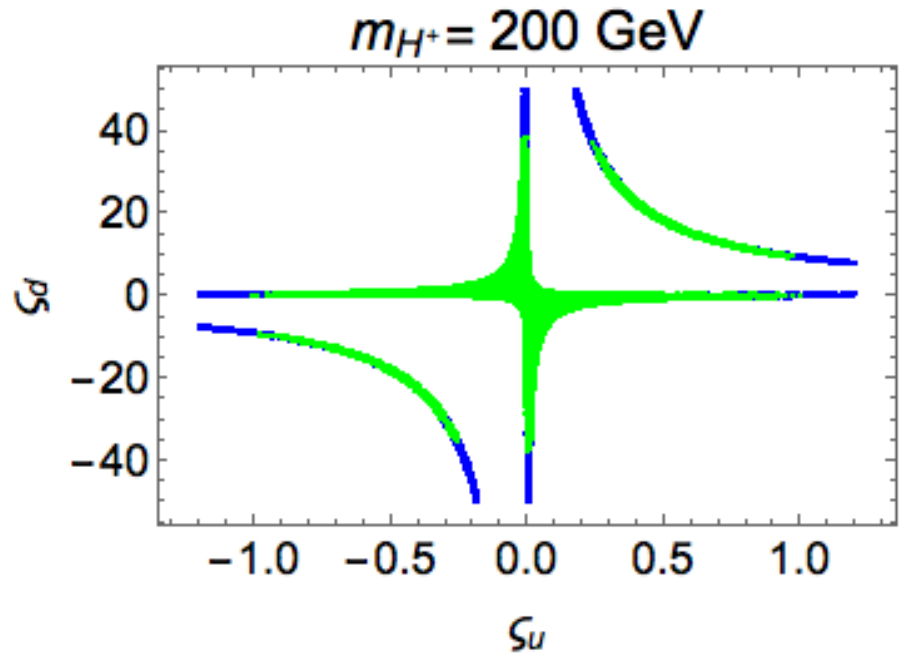}
	\includegraphics[width=7.5cm, height=6cm]{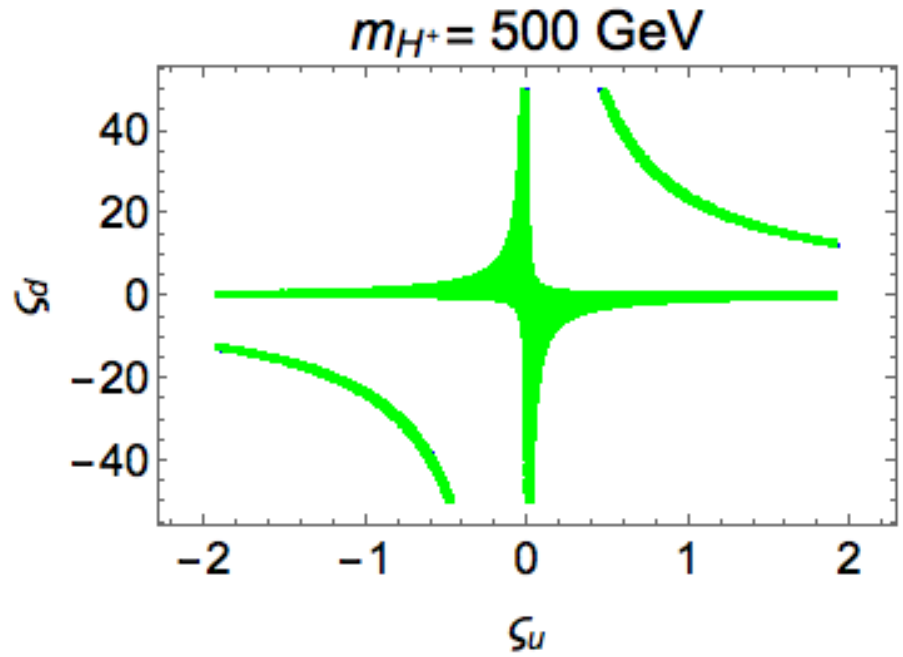}
	\caption{The regions in $\varsigma_u - \varsigma_d$ plane that are allowed by flavor physics data (blue points), and constraints from the $tb$ decay channel of the charged Higgs where $H^\pm$ is produced in association with a top quark at the 13 TeV LHC (green points). The regions are shown for two choices of the charged Higgs mass $m_\Hpm$ = 200 and 500 GeV.}
	\label{fig:lhc_bound1}
\end{figure}

\begin{figure}[htb]
	\centering
	\includegraphics[width=7.5cm, height=6cm]{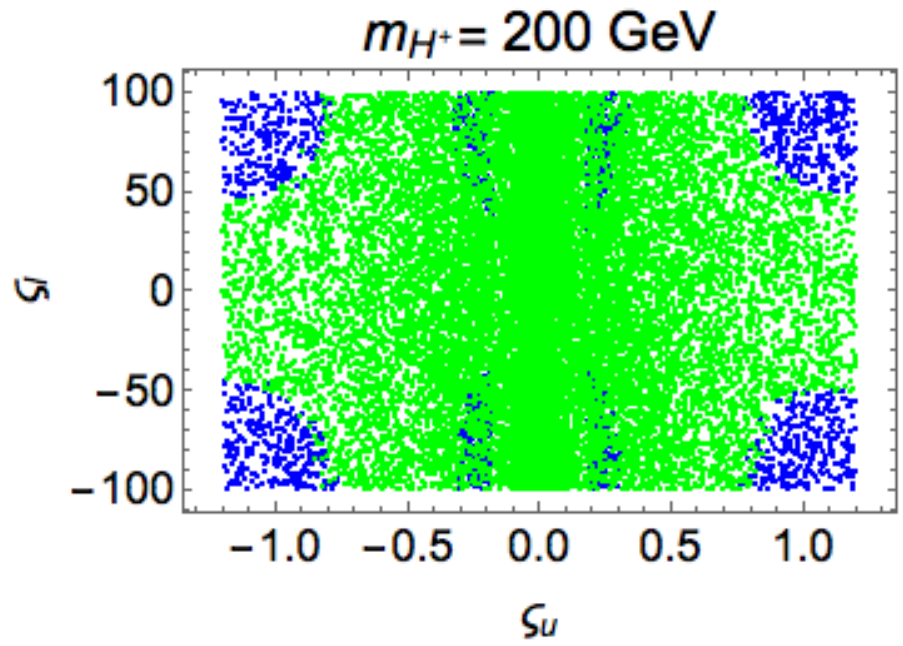}
	\includegraphics[width=7.5cm, height=6cm]{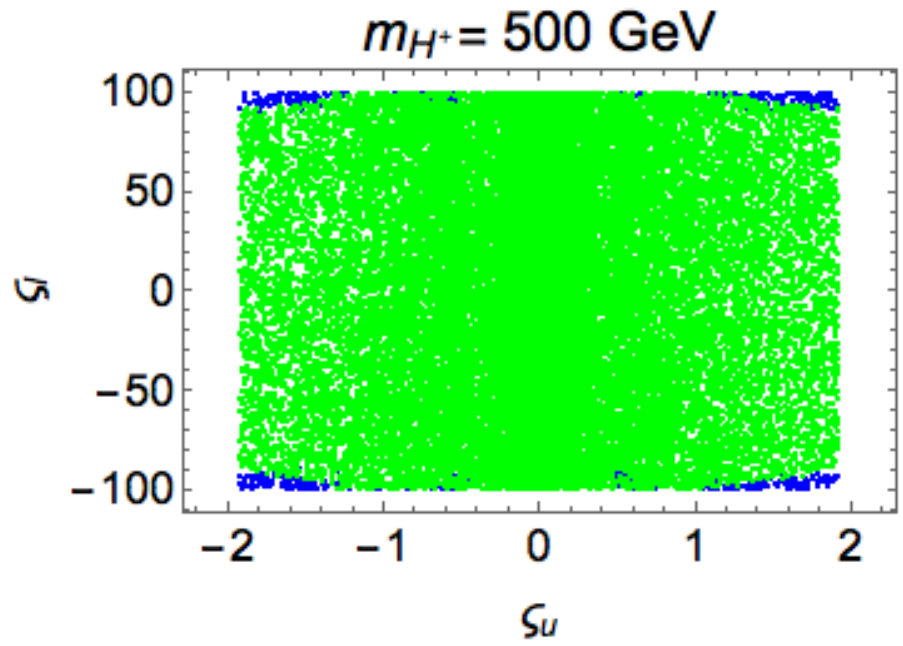}	
	\caption{The regions in $\varsigma_u - \varsigma_\ell$ plane that are allowed by flavor physics data (blue points), and constraints from $\tau \nu_\tau$ decay channel of the charged Higgs where $H^\pm$ is produced in association with a top quark at the 13 TeV LHC (green points). The regions are shown for two choices of the charged Higgs mass $m_\Hpm$ = 200 and 500 GeV.}
	\label{fig:lhc_bound2}
\end{figure}

Following these discussions, we now present in Figs.~\ref{fig:lhc_bound1} and \ref{fig:lhc_bound2} the constraints on the $\varsigma_u - \varsigma_d$ and $\varsigma_u - \varsigma_\ell$ parameter space from the LHC process $\sigma(pp \rightarrow t(b)H^\pm) \times BR(H^\pm \rightarrow \tau^\pm \nu_\tau,tb)$. To obtain the bounds we simply demand that the theoretical value of the quantity $\sigma(pp \rightarrow t(b)H^\pm) \times BR(H^\pm \rightarrow \tau^\pm \nu_\tau,tb)$ is smaller than the LHC limit.  For simplicity the results are shown for two choices of the charged Higgs mass, 200 and 500 GeV and unless otherwise mentioned this choice will be used in the numerical analyses presented in this work.  The blue region is the one allowed by the flavour observables, whereas the region allowed when including the LHC information is shown in green. Overall, the green region is allowed by both the LHC and the flavour physics constraints.  It can be seen from the left plot of  Figs.~\ref{fig:lhc_bound1}, \ref{fig:lhc_bound2}  that only for low charged Higgs masses ($m_{H^+} <$ 500 GeV) the LHC search is currently sensitive to the alignment parameter space allowed by flavour physics. The allowed range of the aligned parameter $\varsigma_d$  for $m_{H^+} =200$ GeV lies between approximately $-40$ and +40, and of the parameter $\varsigma_l$ is between approximately $-50$ and +50 as can be observed from Figs.~\ref{fig:lhc_bound1}, \ref{fig:lhc_bound2}.
%%%%%%%%%%%%%%%%%%%%%%%%%%%%%%%%%%%%%%%%%%%%%%%%%%%%%%%%%%%%%%%%%%%%%%%%%%%%%%%%%%%%%%%%%%%%%
%%%%%%%%%%%%%%%%%%%%%%%%%%%%%%%%%%%%%%%%%%%%%%%%%%%%%%%%%%%%%%%%%%%%%%%%%%%%%%%%%%%%%%%%%%%%%
\section{The $H^\pm \rightarrow  W^\pm Z$ decay in the A2HDM}
\label{sec4}
In this section, we compute the $H^\pm W^\mp Z$ vertex at one-loop in the A2HDM. We have performed the calculations analytically, and to reduce any risk of errors, our computations are tested by specific one-loop open source packages.  The \texttt{FeynCalc} package \cite{Mertig:1990an,Shtabovenko:2016sxi} is used in the analytical computations. The  packages which are being used to test our analytical results are  the publicly available \texttt{FeynRules}~\cite{Alloul:2013bka} model files for 2HDM in which we have implemented the A2HDM.  We generate the \texttt{FeynArts}~\cite{Hahn:2000kx} model files in \texttt{FeynRules} and the amplitudes are calculated using \texttt{FormCalc}~\cite{Hahn:1998yk}.  We have also compared our results numerically using \texttt{LoopTools}~\cite{Hahn:1998yk}.  The diagrams are calculated here in the 't Hooft-Feynman Gauge.
 \begin{figure}[htb!]
\centering
\includegraphics[width=11cm, height=10cm]{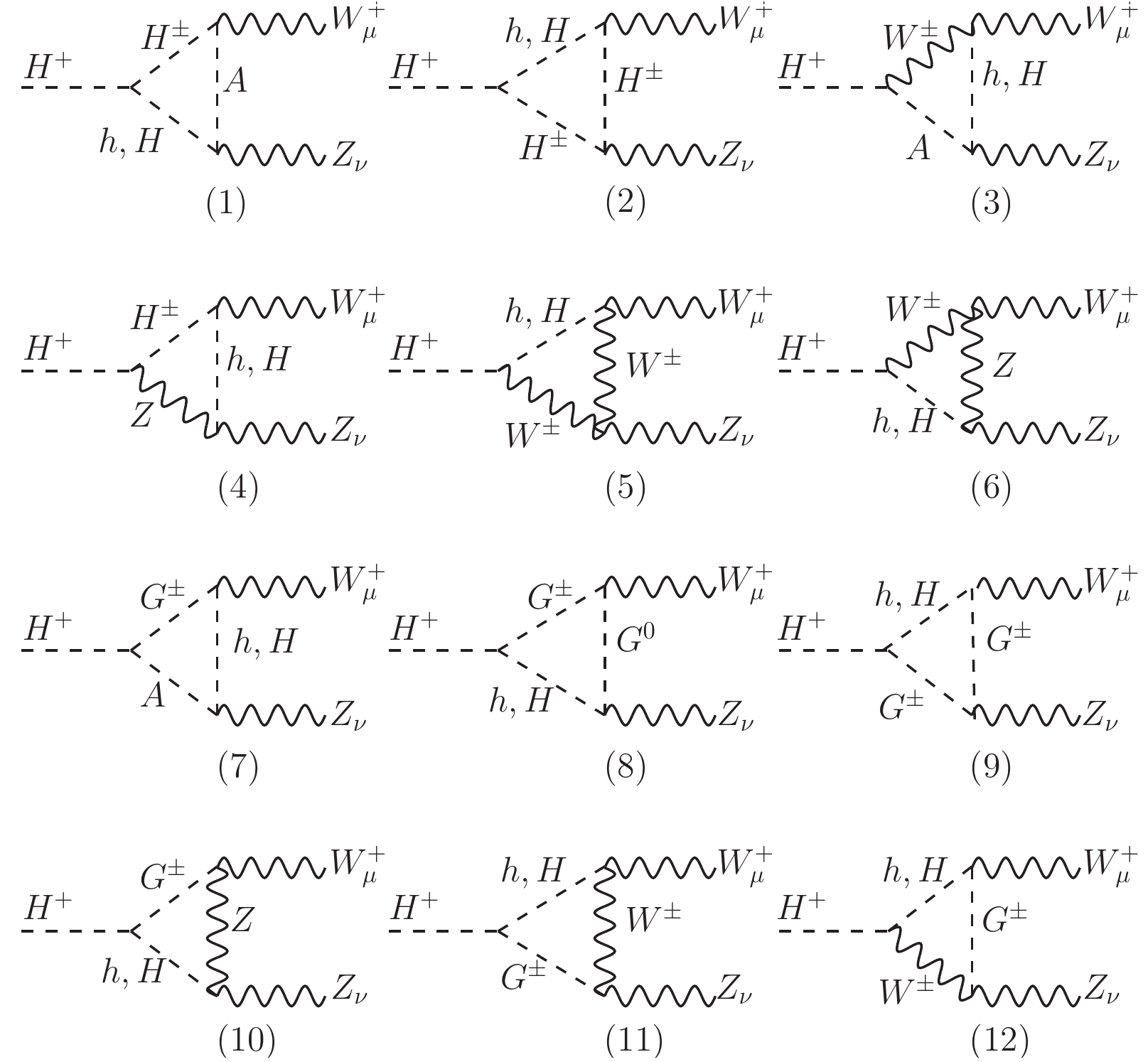}
\caption{The boson-loop triangle diagrams for $H^+\rightarrow W^+Z$ in the 't Hooft-Feynman gauge }
\label{fig1}
\end{figure}

\begin{figure}[htb!]
\centering
\includegraphics[width=11cm, height=8cm]{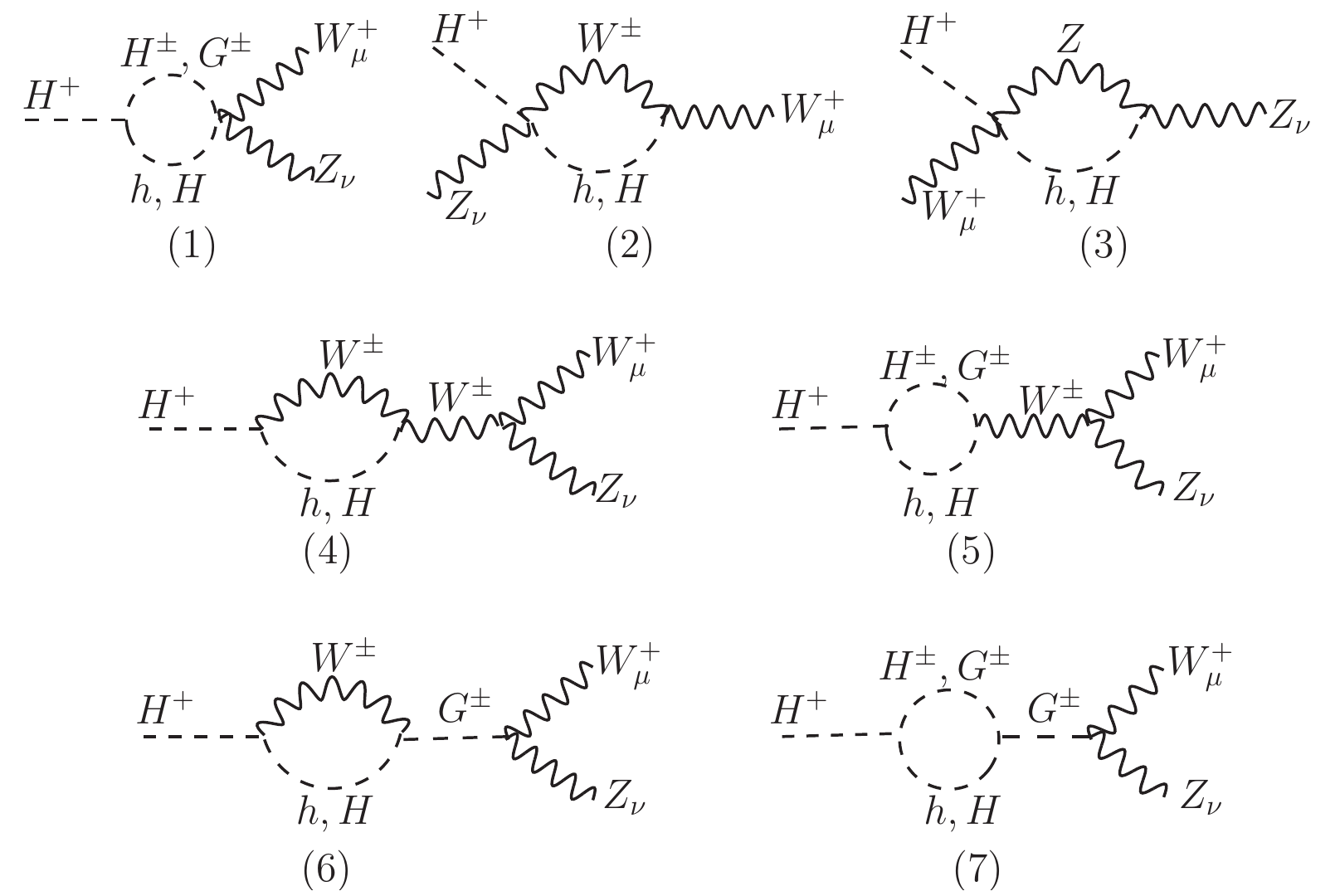}
\includegraphics[width=12cm, height=4.5cm]{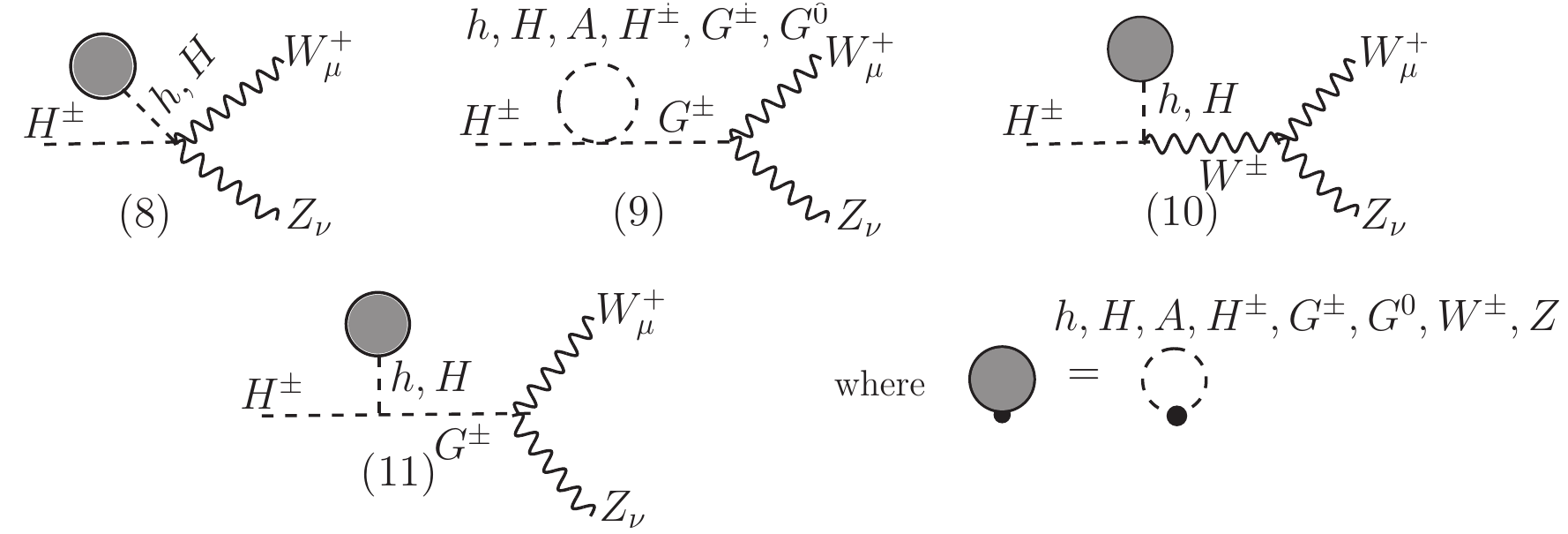}
\caption{The boson-loop and tadpole diagrams for $H^+\rightarrow W^+Z$ in the 't Hooft-Feynman gauge }
\label{fig2}
\end{figure}

\begin{figure}[htb!]
\centering
\includegraphics[width=12cm, height=4cm]{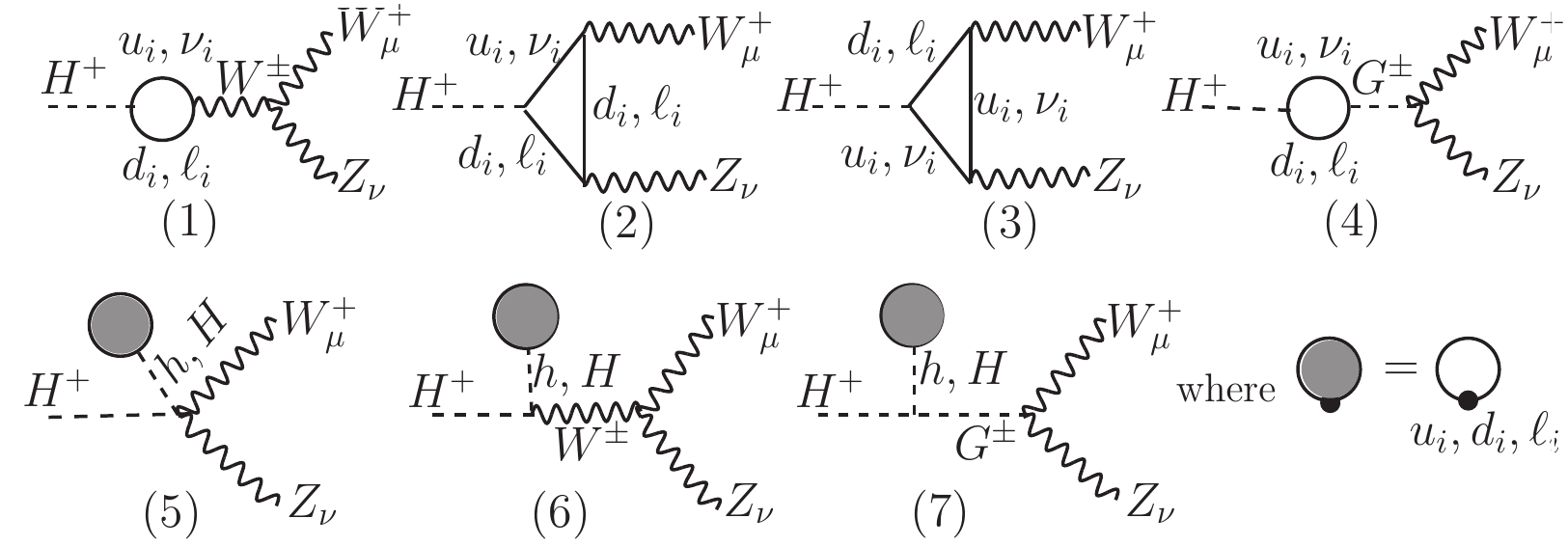}
\caption{The fermion-loop diagrams for $H^+\rightarrow W^+Z$ in the 't Hooft-Feynman gauge. Note that $i$ in $u_i,d_i,\ell_i,\nu_i$ stands for the fermion generation.  }
\label{fig3}
\end{figure}

The loop contributions of the scalars/bosons to the $H^\pm W^{\mp}Z$ vertex are shown in Figs.~\ref{fig1},~\ref{fig2} and the contributions of fermion loops are shown in Fig.~\ref{fig3}. We have  parametrized the $H^\pm \rightarrow  W^\pm Z$ amplitude as
\begin{eqnarray}\label{eq:mat_element}
\mathcal{M}&=&g m_W \mathcal{M}_{\mu\nu}\epsilon_W^{\mu\ast}\epsilon_Z^{\nu\ast},\quad \mathrm{with}\nonumber \\
\mathcal{M}_{\mu\nu}&=& \mathcal{F} g_{\mu\nu}+\frac{\mathcal{G}}{m_W^2} p_{Z\mu}p_{W\nu}+\frac{\mathcal{H}}{m_W^2}\epsilon_{\mu\nu\rho\sigma}p_Z^\rho p_W^\sigma,
\end{eqnarray}
where $\epsilon_{W,Z}^{\mu}$ are the polarizations of the gauge bosons and $p_{W,Z}$ are the momenta. The decay width for $H^\pm \rightarrow W^\pm Z$ in terms of the form factors $\mathcal{F}$, $\mathcal{G}$ and $\mathcal{H}$ listed in Eq.~(\ref{eq:mat_element}) is given as 
\begin{eqnarray}
  \Gamma (H^+ \rightarrow W^+ Z)
    = 
 m_{H^\pm} \fr{\lambda^{1/2}(1,w,z)}{16 \pi} 
  \lt( | \mathcal{M}_{LL} |^2 + | \mathcal{M}_{TT}|^2 \rt),  \label{width}
\end{eqnarray}
where  $w = m_W^2/m^2_{H^\pm}$, $z = m_Z^2/m^2_{H^\pm}$ and 
 $\lambda (a,b,c) = (a - b - c)^2 - 4 b c$. The amplitudes ${\cal M}_{LL}$ and ${\cal M}_{TT}$ contain the contributions from the longitudinally and transversely polarized gauge bosons and are given by
\begin{align}
&  |{\cal M}_{LL}|^2 = \fr{g^2}{4z}
    \lt| (1 - w - z) \mathcal{F} +\fr{\lambda (1,w,z) }{2w}   \mathcal{G} \rt|^2 \label{ll}\\
 & |{\cal M}_{TT}|^2 = g^2\left(2 w |\mathcal{F} |^2 + \fr{\lambda (1,w,z)}{2w} \big|\mathcal{H} \big|^2 \right) \label{tt} , \quad \mathrm {with} \nonumber \\
  & \mathcal{F}    = \frac{1}{g m_W} F, \quad \mathcal{G}  = \frac{m_W}{g} G,  \quad \mathcal{H}  = \frac{m_W}{g} H 
  \end{align}
The contributions to $F,~G,~H$ from individual diagrams in Figs.~\ref{fig1},~\ref{fig2} and \ref{fig3} are listed in tables \ref{tab:coupling1}, \ref{tab:coupling2}  and \ref{tab:coupling3}  of Appendix \ref{sec:FGHterms}. The $\mathcal{F}$ term receives contributions from all the diagrams of Figs.~\ref{fig1},~\ref{fig2},~\ref{fig3} whereas $\mathcal{G}$ only receives contributions from the boson and fermion triangle diagrams.  The fermion loop triangle diagrams only contribute to $\mathcal{H}$ as the boson sector in our case has the parity symmetry.

The dominant contributions to the $H^\pm W^\mp Z$ vertex come from the top quark mass as well as from the non-decoupling effects of the masses of the heavy scalars running in the loop. In the context of the Type II 2HDM, it was discussed in Ref.~\cite{Kanemura:1997ej} that, the $H^\pm tb$ coupling being proportional to $m_t \cot\beta$ and $m_b \tan\beta$, the fermion loop contributions rapidly decrease for larger $\tan\beta$. The decrease of the fermion loop contributions in the case of large $\tan\beta$ is compensated by the scalar non-decoupling effects, with a large mass splitting between $m_A$ and $m_\Hpm$.  Overall, the decay width in the 2HDM is proportional to the top quark contribution in the low $\tan\beta$ region, and to the scalar non-decoupling effects in the large $\tan\beta$ region.  

In the A2HDM, the dominant fermionic contributions to the $H^\pm W^\mp Z$ vertex are proportional to $m_t \varsigma_u$, $m_b \varsigma_d$, and $m_\tau \varsigma_\ell$. Hence, for sufficiently large $\varsigma_u$, the magnitude of the $H^\pm W^\mp Z$ vertex could be enhanced even for small values of the aligned parameter $\varsigma_d$. This is starkly different from the results in the 2HDM of type \rom{2}. The boson loop contributions to the $H^\pm W^\mp Z$ vertex are mainly dependent on the splitting of the charged and pseudoscalar Higgs masses and the three independent parameters of the scalar potential $\lambda_{2,3,7}$.

Since there are too many free parameters involved, we will show our results for some particular benchmark values.  Our results do not deviate drastically if we change these benchmark values.
\begin{figure}[htb!]
\includegraphics[width=7.5cm, height=6cm]{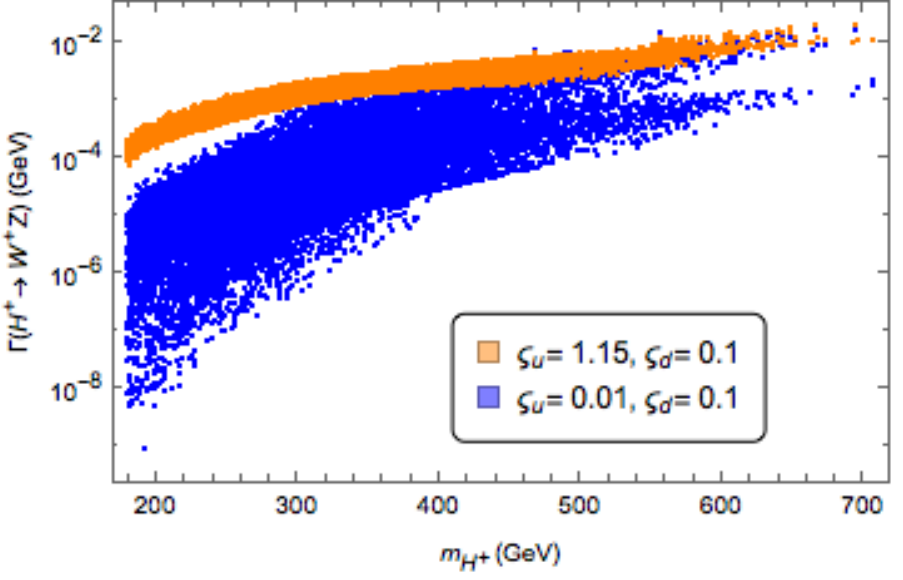}
\includegraphics[width=7.5cm, height=6cm]{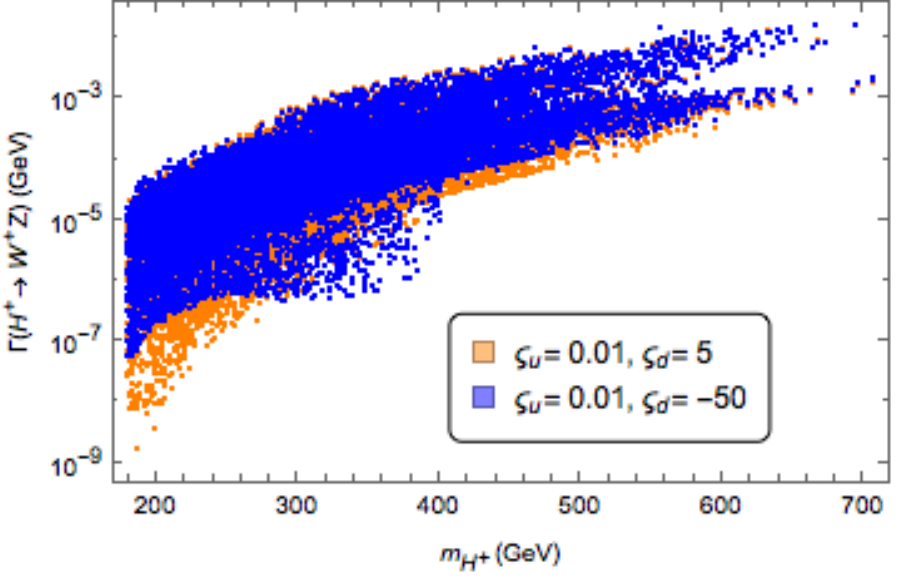}
\caption{The decay width $\Gamma(H^+\rightarrow W^+Z)$ as a function of $m_H^+$ for various values of $\varsigma_u$ and $\varsigma_d$ with $\cos\tilde{\alpha}$ = 0.95,  $|m_\Hpm - m_A| > $ 200 GeV and $m_H = m_\Hpm \pm 15$ GeV. The couplings $\lambda_{3,7,8}$ take the values allowed by the theoretical constraints whereas the alignment parameter $\varsigma_\ell=50$.}
\label{fig:DW2}
\end{figure}
We first explore the dependence of the decay width on the charged Higgs mass and show in Fig.~\ref{fig:DW2} the decay width as a function of $m_\Hpm$ for the mass splittings $|m_\Hpm - m_A| >$ 200 GeV and $m_H=m_\Hpm \pm15$ GeV. The couplings $\lambda_{2,3,7}$ are varied in the allowed range  satisfying the required experimental and theoretical constraints as discussed before. We have fixed the mixing angle value to  $\cos\tilde{\alpha}$ = 0.95, in accordance with the latest LHC results for all our calculations unless otherwise mentioned. 
The left plot shows the variation for two choices of $\varsigma_u$=0.01 and 1.15 and $\varsigma_d$ is fixed to 0.1. In the plot to the right we fix $\varsigma_u$ at 0.01 and $\varsigma_d$ is chosen 5 and -50.  We have explicitly checked that the decay width does not change much with $\varsigma_\ell$, therefore it is kept fixed at 50 for these plots.  These figures show that the decay widths are quite sensitive to $\varsigma_u$ for a given charged Higgs mass.  

We next consider the Higgs effect from the scalar loop diagrams and show the dependence of the decay width on the Higgs mass splittings and the $\lambda_{2,3,7}$ parameters. The bosonic-diagrams with the SM-like Higgs $h$ in the loop are proportional to $\sin\tilde{\alpha}$ and therefore have a very small contribution. Therefore, a large value of the scalar self-coupling constant $\lambda_7$ (proportional to $\cos\tilde\alpha$ for the diagrams with $H$ in the loop) is considered in order to make the contributions from the boson and fermion loop diagrams comparable.  The parameter $\lambda_2$ does not contribute to our process, whereas $\lambda_3$ is always accompanied with $\sin\tilde\alpha$ in most of the diagrams that contribute. We therefore work in the limit where the parameters $\lambda_{2,3}$ are fixed to zero and a large non-zero value for $\lambda_7 \simeq$ 8 is considered.
\begin{figure}[htb!]
\includegraphics[width=7.7cm, height=6cm]{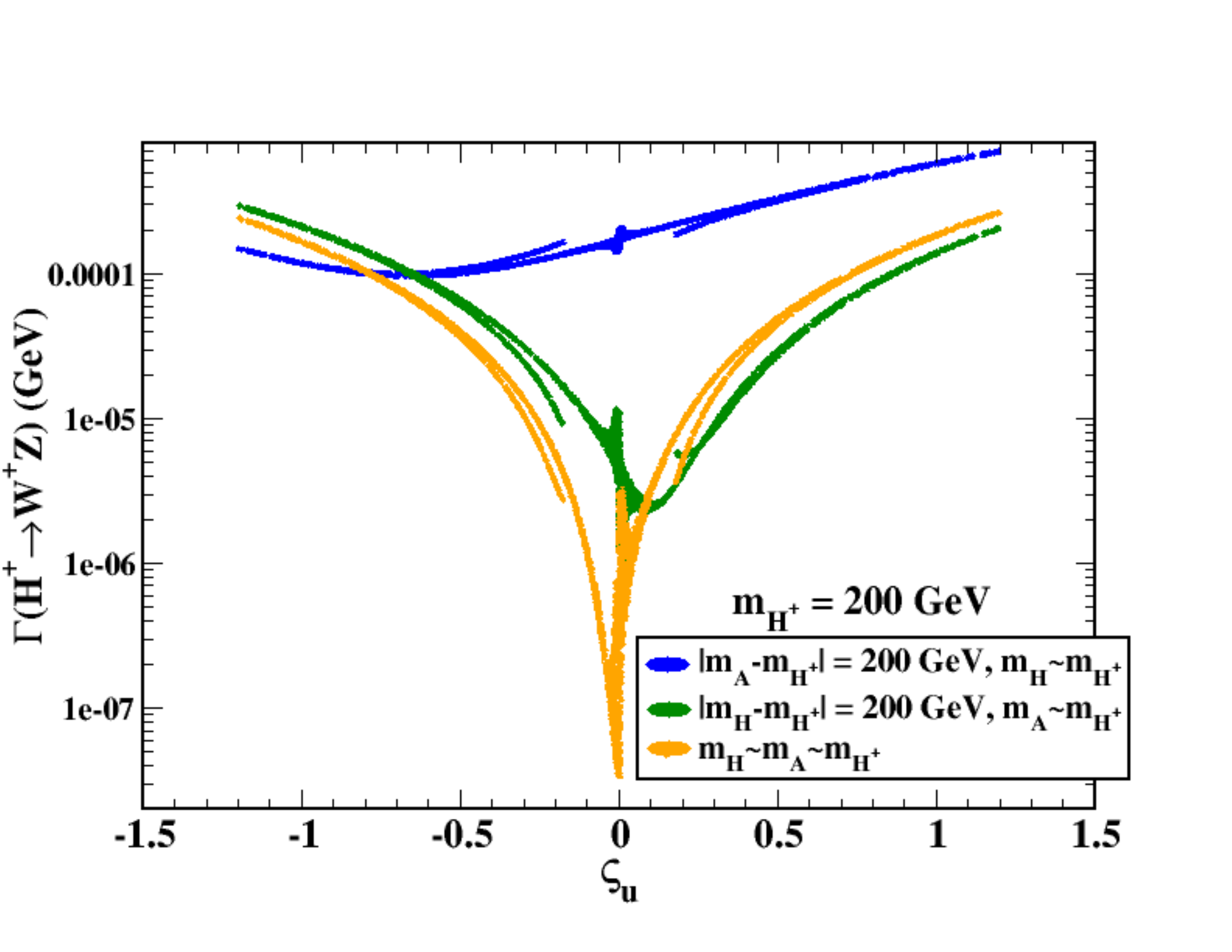}
\includegraphics[width=7.7cm, height=6cm]{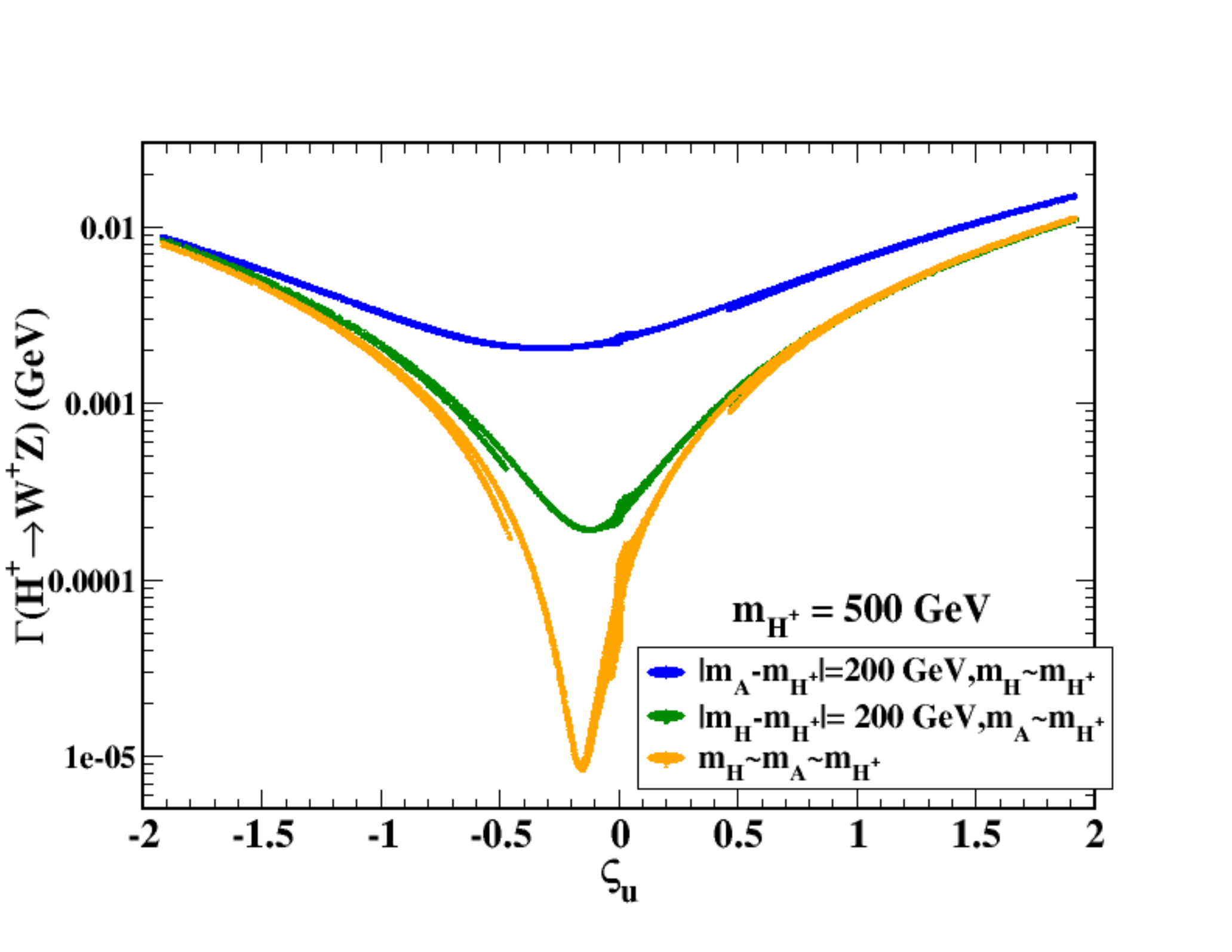}
\caption{Decay width $\Gamma(H^+\rightarrow W^+Z)$ as a function of $\varsigma_u$ for different mass splitting between charged and CP-odd scalars with  $\cos\tilde{\alpha}$ = 0.95. The parameter $\varsigma_d$ is varied in the allowed range, whereas the other parameters are fixed as discussed in the text. The figures are shown for two choices of charged Higgs mass $m_{H^+}$ = 200 and 500 GeV.}
\label{fig:DW3}
\end{figure}

We now show in Fig.~\ref{fig:DW3} the $\varsigma_u$ dependence of the decay width for $m_{H^+}$ = 200, 500 GeV with various mass splittings between the charged and the CP-odd scalars. The $\lambda_{2,3,7}$ parameters are fixed to values as discussed above and $\varsigma_d$ is varied in the range [-50,50].  The decay width dependence on the different mass splitting scenarios can be easily interpreted from the figure. The case where the additional scalars are degenerate (orange points) is sensitive to $\varsigma_u$, as the contribution from scalar-loop diagrams gets suppressed with respect to the remaining contributions from the fermions and the gauge bosons loop diagrams. The top mass contribution to the decay width becomes very small in the limit where $\varsigma_u$ tends to zero. The blue points in Fig.~\ref{fig:DW3} with a mass splitting $m_A-m_{H^+}$  = 200 GeV show that the decay width is not small for small $\varsigma_u$. This is because the top mass effect is dominant at large $\varsigma_u$, whereas in the low $\varsigma_u$ region the top mass contribution is decreased but the non-decoupling effects of heavier scalars increases the strength of the $H^\pm W^\mp Z$ vertex.

The reason for enhancement for the case where the CP-odd scalar is degenerate with the charged Higgs while there is a large mass splitting between the heavy CP-even scalar and the charged Higgs  (green points) is similar to the orange region, that is, large values of  the aligned parameter  $\varsigma_u$ and the mass of the top quark contribute through the fermionic loop.  The results for the decay width shown in  Fig.~\ref{fig:DW3} are sensitive to $\lambda_7$, as the $HH^+H^-$ vertex in the scalar loop diagrams is proportional to $\lambda_7$. Therefore, with the decrease in the value of $\lambda_7$ the contribution from the fermionic loop diagrams becomes dominant and the decay width becomes sensitive to $\varsigma_u$, irrespective of the mass-splitting between the scalars.  The contribution from the scalar loop diagrams is  therefore dominant with a large mass splitting and large allowed values of $|\lambda_7|$, for $\cos\tilde\alpha$ =0.95.     

\begin{figure}[htb!]
% \hspace{-0.5cm}
\includegraphics[width=7.5cm, height=6cm]{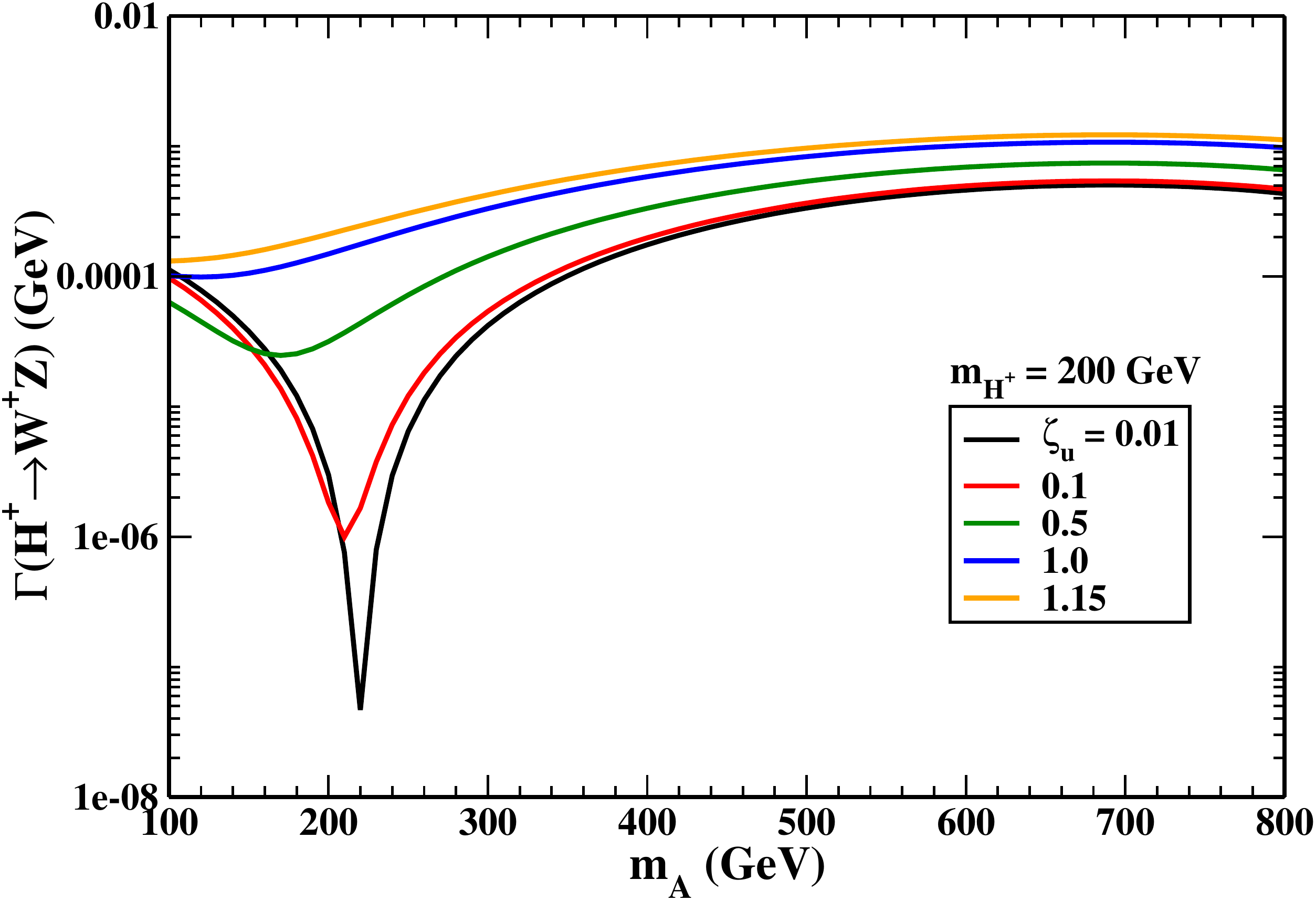}
\includegraphics[width=7.5cm, height=6cm]{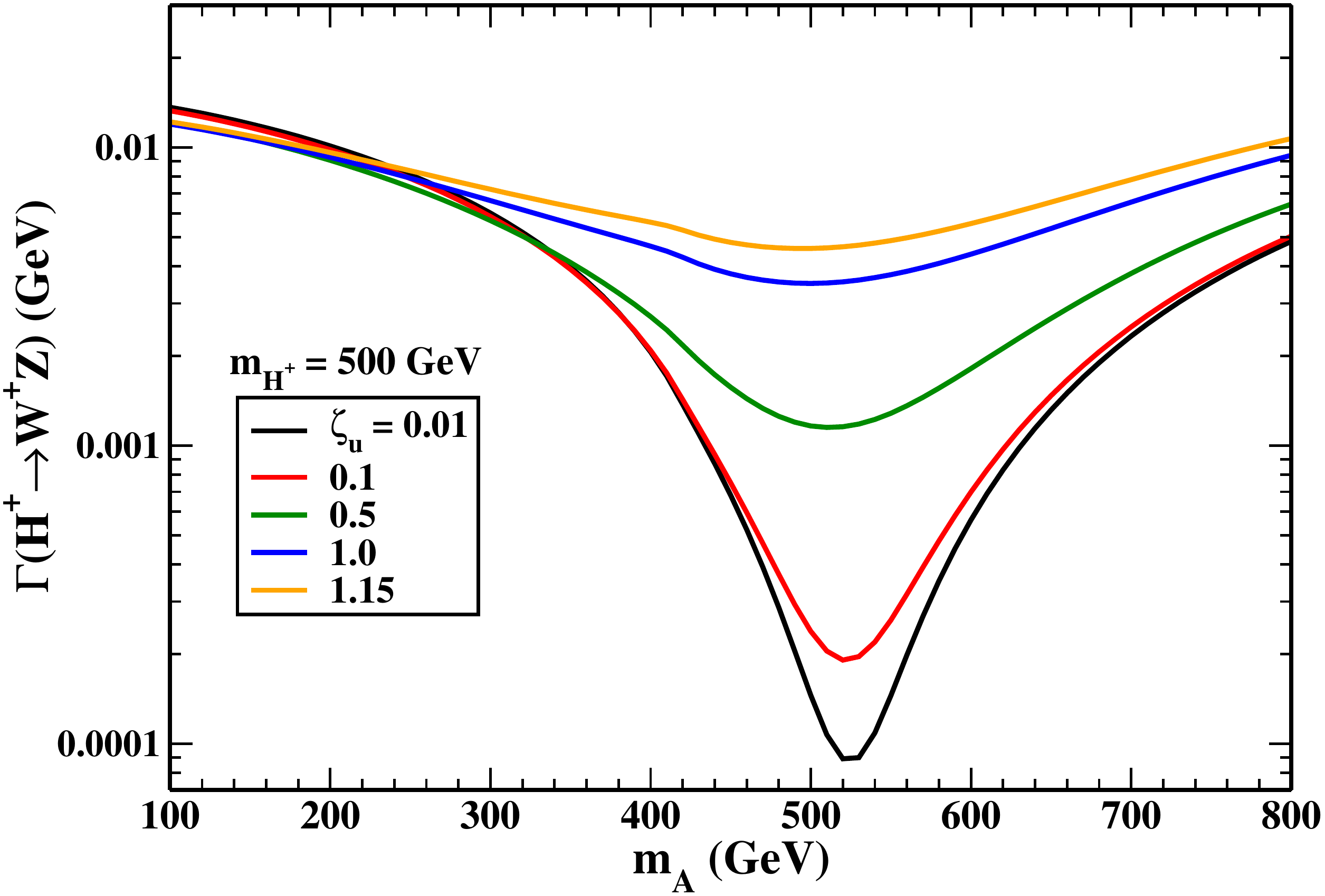}
\caption{The decay width $\Gamma(H^+\rightarrow W^+Z)$ as a function of $m_A$ for  different values of the Yukawa alignment parameter $\varsigma_u$ with $\varsigma_d$ = 0.1, $\varsigma_\ell$ = 50, $\cos\tilde{\alpha}$ = 0.95, $m_H = m_{H^+} +10$ GeV, and $\lambda_{2,3}=0$. The figures are shown for two values of charged Higgs mass $m_{H^+}=200, 500$ GeV.}
\label{fig:DW4}
\end{figure}

The Higgs mass effect is dominant when the $m_A-m_{H^+}$ mass splitting is large. We discuss this scenario in details in the following. We plot the decay width as a function of $m_A$ for different choices of $\varsigma_u$ in Fig.~\ref{fig:DW4}.  The results remain the same with the variation of $\varsigma_d$ and $\varsigma_\ell$, which we fix at 0.1 and 50 in these figures. The other parameters are similar to the previous figure. We can see from Fig.~\ref{fig:DW4} that the decay width becomes independent of $\varsigma_u$ for large mass splitting between $m_A$ and $m_{H^+}$. In the near custodial symmetry limit ($m_A \simeq m_{H^+}$),  the non-decoupling effects of the scalar masses  are highly suppressed and the decay width receives contribution only from the fermionic and the gauge boson diagrams making the decay width sensitive to  $\varsigma_u$.   

We now consider the branching ratio (BR) for the process $H^+ \rightarrow W^+Z$.  The decay is kinematically allowed when the charged Higgs mass $m_\Hpm > m_W+m_Z$. The threshold of $H^\pm \to tb$ is also very close to $m_W+m_Z$.  The $tb$ mode becomes dominant for large values of $\varsigma_{u,d}$ when $m_\Hpm > m_t+m_b$. We note that for $m_\Hpm$ around 200 GeV, only the decay mode $\tau\nu_\tau$ dominates over the $W^+Z$ decay for large values of $\varsigma_\ell$. 
\begin{figure}[htb!]
\includegraphics[width=5.5cm, height=5cm]{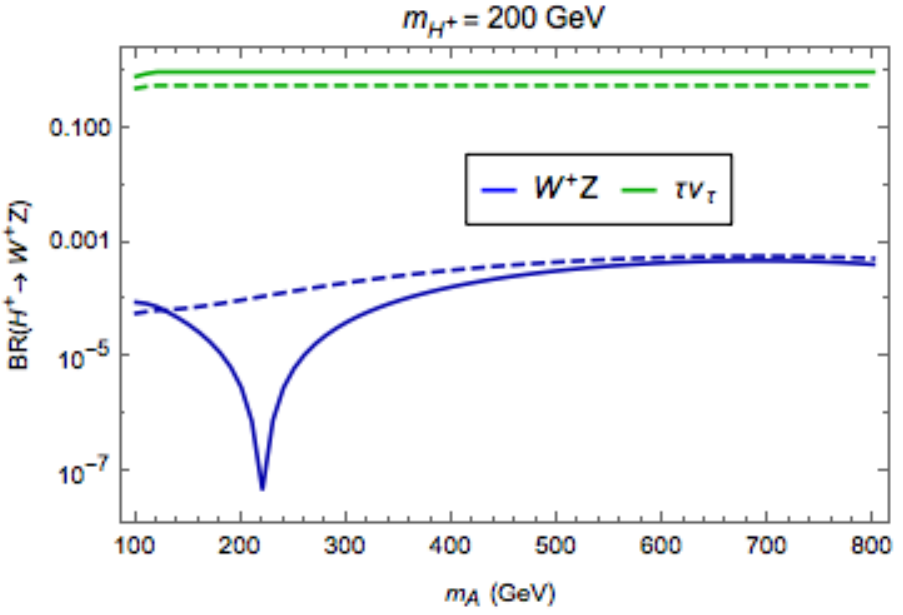}
\includegraphics[width=5.5cm, height=5cm]{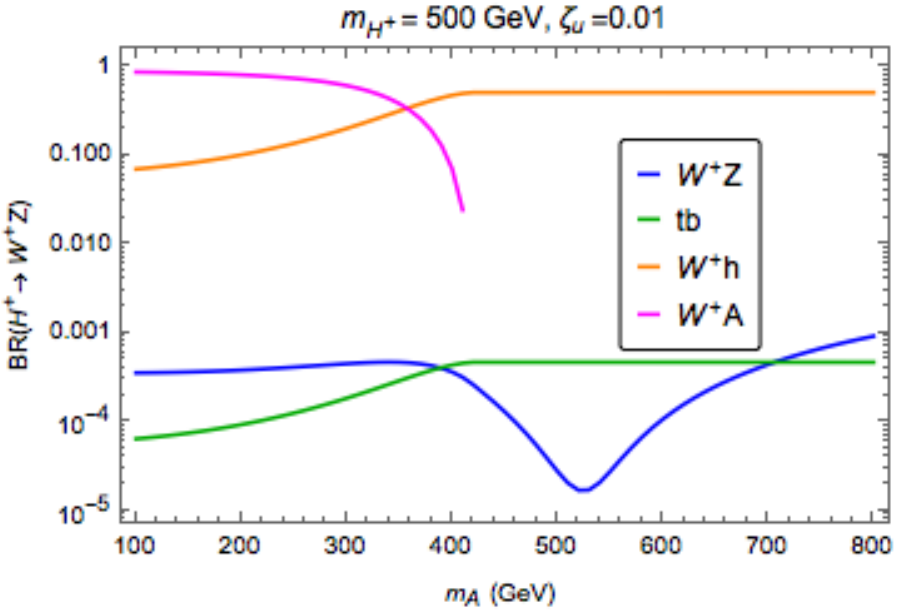}
\includegraphics[width=5.5cm, height=5cm]{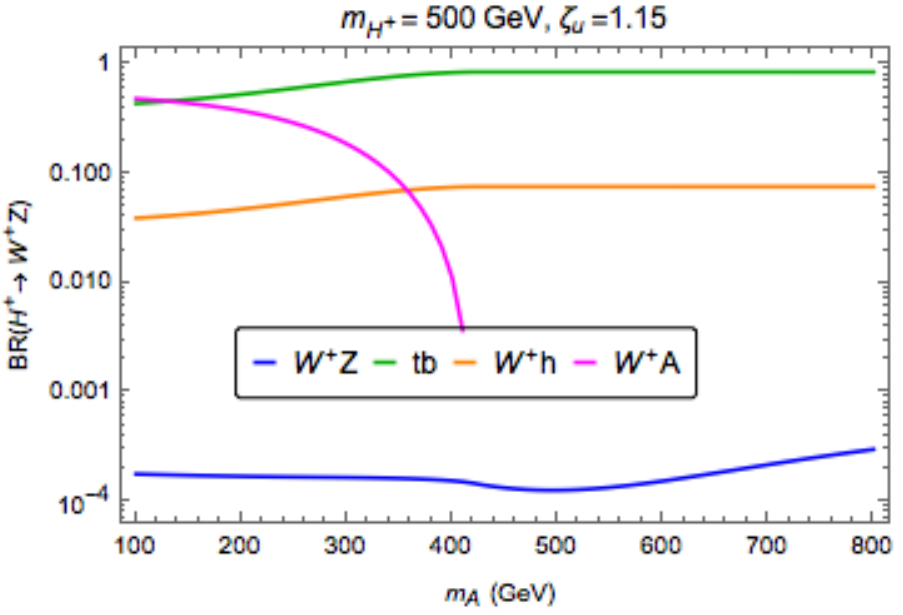}
\caption{The branching ratio $\Gamma(H^+\rightarrow W^+Z)$ as a function of $m_A$ for  $\lambda_{2,3}$ = 0,  $\varsigma_d$ = 0.1, $\varsigma_\ell$ = 50, $m_H = m_{H^+} +10$ GeV,  and $\cos\tilde{\alpha}$ = 0.95. The other dominant branching ratios of $H^+$ are also shown. In the left plot, the solid lines are for $\varsigma_u=0.01$ and the dashed lines for $\varsigma_u=1.15$.}
\label{fig:BR1}
\end{figure}
This can be seen from the first plot of Fig.~\ref{fig:BR1} where we show the branching ratio of the charged Higgs as a function of $m_A$. The $\tau\nu_\tau$ mode is shown with green color and the $W^+Z$ mode in blue. With the $\varsigma_\ell=50$, the leptonic decay channel has a branching ratio of almost 1, whereas the BR of  $W^+Z$ increases with the variation of $\varsigma_u=0.01$ (solid) to $\varsigma_u=1.15$ (dashed). Here the values of $\lambda_{2,3}$ are fixed at zero, $\lambda_7 \simeq$ at 8 and $\varsigma_d$ at 0.1. In the limit of vanishing $\varsigma_\ell$ and $m_\Hpm$ around 200 GeV, the BR for $W^+Z$ will be 1. For a heavy charged Higgs, various other decay channels open up as can be seen from the second and third plot of Fig.~\ref{fig:BR1}. The $tb$ final state has a dominant branching ratio for large values of $\varsigma_u$.  The BR($H^+ \rightarrow W^+ Z$)  is larger than the one into $tb$ when the alignment parameters are small for the region where there is a large mass difference between $m_A$ and $m_{H^+}$, as seen from the second plot of Fig.~\ref{fig:BR1}. With smaller $\varsigma_u$, a larger value of $\varsigma_d$ will lead to an enhanced branching ratio of $tb$ whereas the $W^+Z$  will not  be affected significantly. 

\begin{figure}[htb!]
\includegraphics[width=7.5cm, height=6cm]{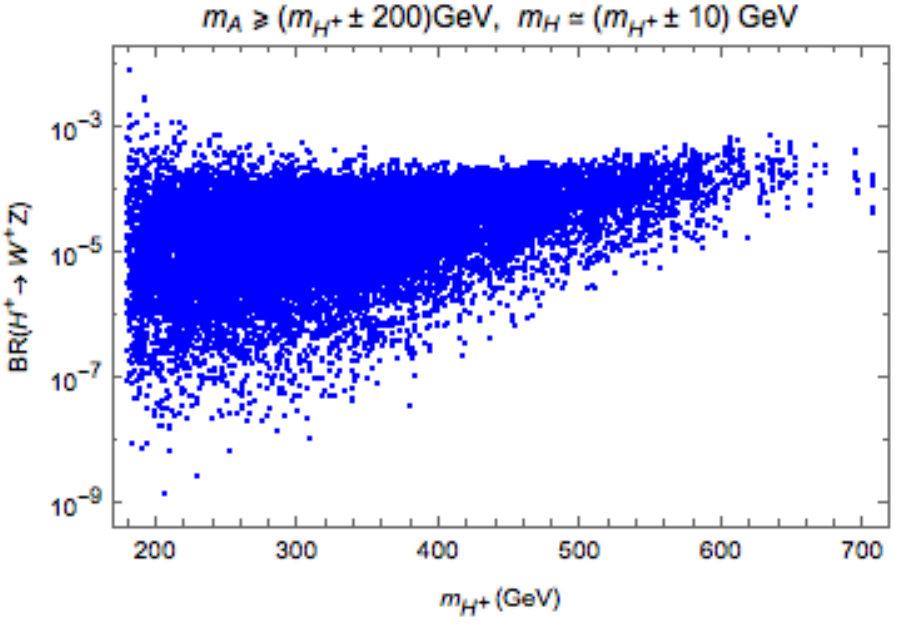}
\includegraphics[width=7.5cm, height=6cm]{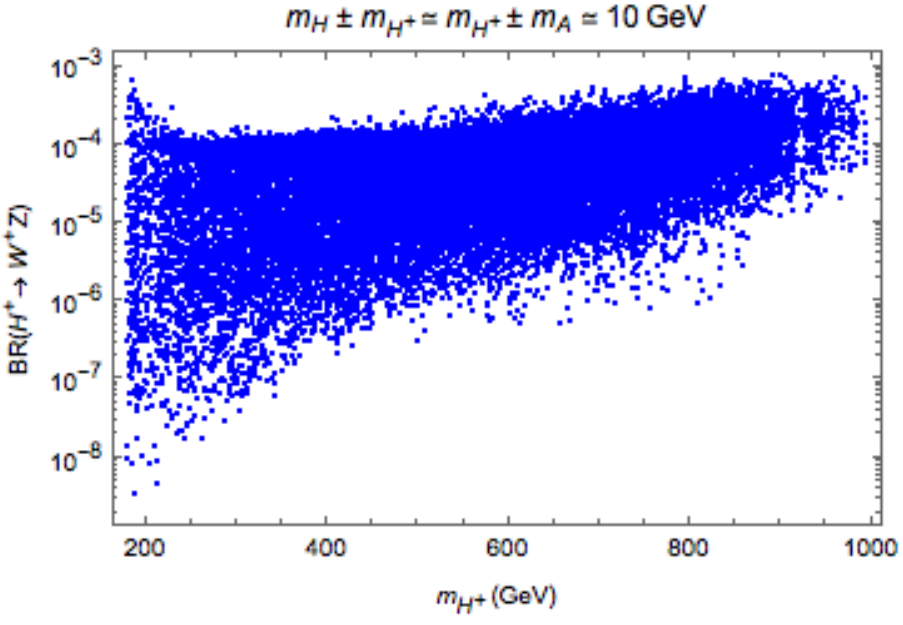}
\caption{The branching ratio $BR(H^+\rightarrow W^+Z)$ as a function of $m_H^+$ for values of $\varsigma_u$,  $\varsigma_d$ and $\varsigma_\ell$ and the $\lambda_{3,7}$ parameters  allowed by the theoretical and the experimental constraints discussed in the text. }
\label{fig:BR3}
\end{figure}

Finally in Fig.~\ref{fig:BR3} we show the branching ratio as a function of $m_\Hpm$ for two different scenarios. The left plot shows the branching ratio as a function of $m_\Hpm$ for the case when $m_A \geq m_\Hpm \pm 200$ GeV and $m_H \approx m_\Hpm \pm 10$ GeV. In the right plot we show the branching ratio for the case  where the CP-odd scalar and the CP-even heavy scalar are  degenerate in mass ($m_A \approx m_H $). In these plots $\lambda_{2,3,7}$ and the alignment parameters are all varied within the region allowed by the theoretical and experimental constraints. The branching ratio in the low mass range can be as large as $10^{-3}$.  A large branching ratio can be obtained for small alignment parameters, large $\lambda_7$ and a large mass difference  $|m_A-m_{H^+}|$. The decay width increases with large $\varsigma_u$, but this also leads to the enhancement of the dominant decay channel $tb$. 

\section{$H^\pm$ production through $WZ$ fusion at the LHC} \label{sec5}
In this section we explore whether the BR($H^\pm \rightarrow W^\pm Z$) in the A2HDM could be  large enough to be detected at the LHC.  The charged Higgs for the mass range considered here will be mainly produced through the process $ p p \rightarrow t[b]H^\pm$  with the dominant decay mode of $H^\pm$ being $H^\pm\to tb$, if kinematically allowed. The cross-section for the $gb \rightarrow t H^\pm$ sub-process at the 13 TeV LHC with $\varsigma_u =1.15,~\varsigma_d = 0.1$ will be $\approx$ 4367 fb for $m_\Hpm$ = 200 GeV and $\approx$ 454 fb for $m_\Hpm$ = 500 GeV. 
This process has been studied at the LHC, and we have discussed the constraints on the alignment parameters from this process and the subsequent decay of the charged Higgs to $tb, \tau\nu_\tau$ in Sec.~\ref{sec3A}.  

\begin{figure}[htbp]
  \begin{minipage}[b]{0.3\linewidth}
    \centering
    \includegraphics[width=5.5cm, height=4.5cm]{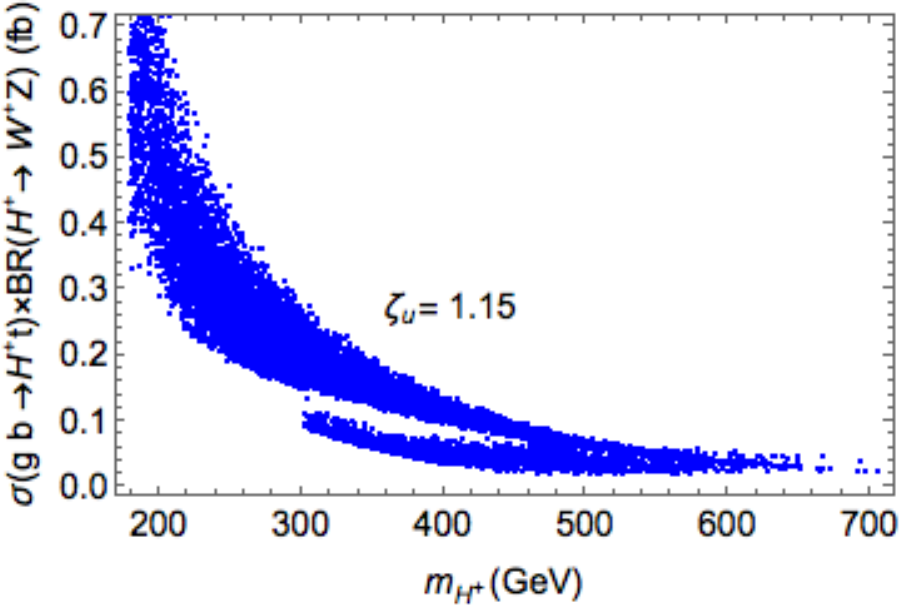}
    \caption{The cross-section of $H^+$ production in association with a top quark followed by its decay to $W^+Z$ as a function of $m_{H^+}$. }
    \label{fig:cs1}
  \end{minipage}
  \hspace{0.5cm}
  \begin{minipage}[b]{0.3\linewidth}
    \centering
    \includegraphics[width=5.5cm, height=4.5cm]{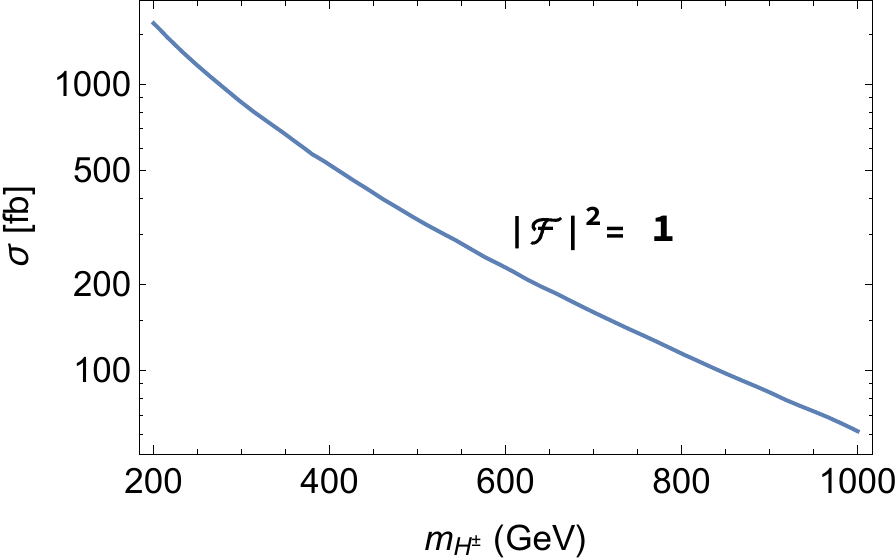}
    \caption{The  cross-section of $H^+$ production through $WZ$ fusion at the 13 TeV LHC, with the form factor $\mathcal{F}$ set to unity and $\mathcal{G} = \mathcal{H}$ = 0. }
    \label{fig:cs2}
  \end{minipage}
  \begin{minipage}[b]{0.3\linewidth}
    \centering
    \includegraphics[width=5.5cm, height=4.5cm]{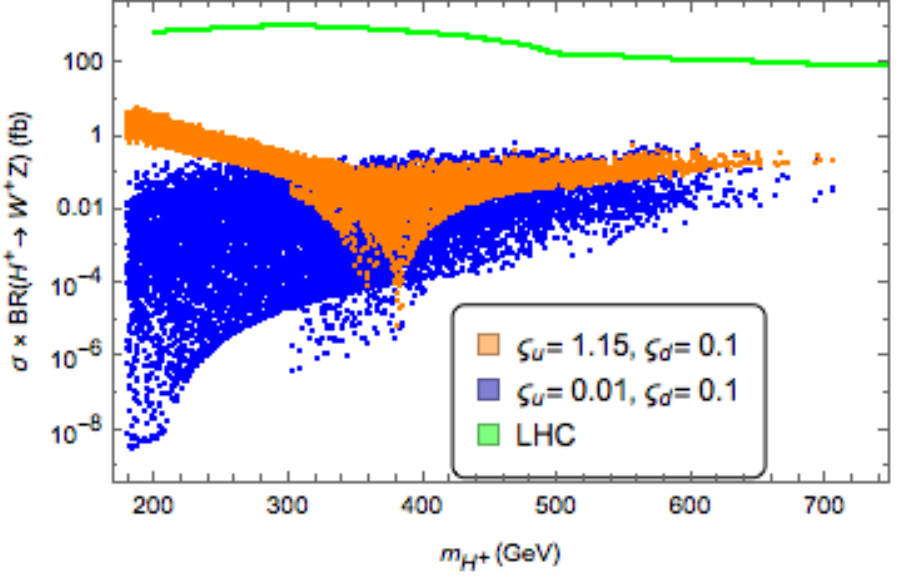}
    \caption{$\sigma$ for $H^\pm$ production through $WZ$ fusion with subsequent decay of  $H^{\pm}$ to $W^{\pm}Z$. The line in green is the current bound from LHC. }
    \label{fig:cs3}
  \end{minipage}
\end{figure}

We show in Fig.~\ref{fig:cs1} the expected cross-section for the process $g b \rightarrow t H^+ \rightarrow t W^+ Z$ at $\sqrt{s}$ = 13 TeV as a function of $m_\Hpm$ with the couplings $\lambda_{2,3,7}$ varied in the allowed range. The alignment parameters are kept fixed as $\varsigma_u =1.15,~\varsigma_d = 0.1$ and we take $|m_A-m_\Hpm|$ = 200 GeV,  $|m_H-m_\Hpm|$ = 10 GeV. The dominant SM background to this process will be $W^+Z+X$, which can be reduced with appropriate kinematic cuts on the final state.   The signal in this final state can be  observed at the high luminosity LHC for low $m_\Hpm $ and $\varsigma_u \gtrsim 1$.  

We next discuss the charged Higgs production through $WZ$ fusion at the 13 TeV LHC, and show in Fig.~\ref{fig:cs2} the expected cross section  assuming $\mathcal{G}=\mathcal{H} $= 0 for simplification. This is a good approximation as the  loop induced $\mathcal{F}$ term is much larger than the $\mathcal{G}$ and  $|\mathcal{H}|$ terms for most of the parameter space.  The analytical form for the cross section is given in Ref.~\cite{Asakawa:2005gv}, however we have performed our computation in \texttt{Madgraph} and compared it with their results.  This process followed by the decay of the charged Higgs to $WZ$ has been studied by the CMS collaboration~\cite{Sirunyan:2017sbn}.  We present our results in Fig.~\ref{fig:cs3} considering the parameter space where a large branching ratio of $H^\pm \rightarrow W^\pm Z$ is observed. The $\lambda_{2,3,7}$ parameters are varied taking into account the theoretical and experimental constraints and the mass differences between the additional Higgs boson are fixed at $|m_A-m_\Hpm|$ = 200 GeV,  and $|m_H-m_\Hpm|$ = 10 GeV. The green line is the current experimental bound from the LHC.  At large $m_{H^\pm}$, we see that the cross section becomes comparable for $\varsigma_u$ = 0.01 and 1.15. This is because the decay width of the $tb$ decay channel is proportional to $\varsigma_u^2 (m_\Hpm^2-m_t^2)/m_\Hpm$ and for large $m_\Hpm$ will be $\sim \varsigma_u^2 m_\Hpm$.  Therefore at large $m_\Hpm$ the branching ratio of $W^\pm Z$ decreases with large $\varsigma_u$.   

We observe in  Fig.~\ref{fig:cs2} that the cross-section can be as low as 1 fb as the mass of the charged Higgs goes beyond 1 TeV.  Hence, when the luminosity of the LHC will be 300 $\rm{fb}^{-1}$, 300 events with a charged Higgs can be produced for a 1 fb cross-section.  Similar conclusion can be drawn from Fig.~\ref{fig:cs3} for the high luminosity phase of the LHC.  Thus, we see that the  $H^\pm \rightarrow W^\pm Z$ in the A2HDM is within the reach of the high luminosity phase of the LHC \cite{Apollinari:2017cqg,Chiang:2012cn,Zaro:2015ika}.  On the other hand, the $tH^\pm$ production channel will be dominant to produce a sufficient number of charged Higgs only in the lower mass region as can be seen from Fig~\ref{fig:cs1} for a luminosity of 300 $\rm{fb}^{-1}$.  However, as the high luminosity phase will move forward, this conclusion about the $tH^\pm$ production channel will not hold.

\section{Summary}
\label{sec6}
The custodial symmetry of the SM could have interesting implications if there exists an extended scalar sector.  This may give rise to remarkable signatures such as an enhancement of the  $H^\pm W^\pm Z$ vertex. This vertex is absent at tree level in the 2HDM because of the weak isospin symmetry of the kinetic terms of the Higgs sector and appears at one- loop.  Furthermore, this vertex was studied in the 2HDM of type \rom{2} before the discovery of the SM-like Higgs at the LHC \cite{Kanemura:1997ej,Kanemura:1999tg} .

In this work we have investigated the $H^\pm W^\pm Z$ vertex using the 't Hooft Feynman gauge in the A2HDM.  The computation is performed taking into account  the theoretical constraints such as the vacuum stability, perturbative unitarity, as well as the bounds from the electroweak precision data. The experimental bounds from flavour physics observables as well as the direct searches of $H^\pm$ at the LHC are also taken into account. The latest results from the charged Higgs searches at the LHC are used to constrain the alignment parameters, as the production and decay mode of the charged Higgs are proportional to them. We find that for $m_{H^+} <$ 500 GeV  the LHC data from charged Higgs searches constrains the aligned parameters $\varsigma_{u,d,\ell}$  allowed by the flavour observables. The parameter space for $m_\Hpm \geq$ 500 GeV is currently not sensitive to the LHC results. 

We later discuss the non-decoupling effects of heavy scalars and fermions in the decay width of $H^+ \rightarrow W^+ Z$, and find that the decay width is more sensitive to $m_t \varsigma_u$ compared to $m_{d,\ell}\varsigma_{d,\ell}$.  We note that in the A2HDM for $\cos\tilde\alpha$ tending to 1, the  non-decoupling effects from the boson loop diagrams are proportional to $\lambda_7$ and a large mass splitting between the CP-odd Higgs and the charged Higgs. Hence, large values of the quartic coupling $\lambda_{7}$  helps in enhancing the magnitude of the $H^\pm W^\pm Z$ vertex.   An enhancement of this vertex also occurs when the alignment parameter $\varsigma_u$ is large even if the alignment parameter $\varsigma_d$ is small.

The dependence of the decay width on each of the independent parameters is discussed individually. We have worked for the charged Higgs in the mass range 200-1000 GeV and find that the maximum obtainable branching ratio for the process considered here in light of the recent experimental constraints is around $\mathcal{O}(10^{-3})$. 

Finally we also calculate the two charged Higgs production modes at the LHC and it's subsequent decay to $W^+Z$. The production modes are (1) $H^+$ produced in association with a top quark (2) $H^+$ produced singly through $WZ$ fusion.  The $W^\pm Z$ final state produced through WZ fusion in the A2HDM is within the reach of the future high luminosity phase of the LHC.

\begin{table}[!h]
\centering
\begin{tabular}{|c|c|c|}
\hline 
Fig.~\ref{fig1}[n] &amplitude & argument \\ \hline
% & $F^{\ref{fig1}[n]}$,$G^{\ref{fig1}[n]}$& \hline
 $F^{\ref{fig1}[1]}$&$-\frac{g^2 v}{c_W}\lambda^{\varphi_i^0}_{H^+H^-}\lambda^{\varphi_i^0}_{AZ} C_{24}$ & $\left[p_W,-p_{H^\pm},m_A,m_{H^\pm},m_{\varphi_i^0}\right]$ \\
$G^{\ref{fig1}[1]}$&$-\frac{g^2 v}{c_W}\lambda^{\varphi_i^0}_{H^+H^-}\lambda^{\varphi_i^0}_{AZ} \left(C_{22}-C_{23}\right)$&$\left[p_W,-p_{H^\pm},m_A,m_{H^\pm},m_{\varphi_i^0}\right]$ \\ \hline
%%%%%%%%%%%%%%%%%%%%%%%%%%%%%%%%%%%%%%%%%%%%%%%%%%%%%%%%%%%%%%%%%%%%%%
 $F^{\ref{fig1}[2]}$&$\frac{g^2 v}{c_W}\left(c_W^2-s_W^2\right)\lambda^{\varphi_i^0}_{H^+H^-}\lambda^{\varphi_i^0}_{W^\pm H^\mp} C_{24}$ & $\left[p_W,-p_{H^\pm},m_{H^\pm},m_{\varphi_i^0},m_{H^\pm}\right]$ \\
$G^{\ref{fig1}[2]}$&$\frac{g^2 v}{c_W}\left(c_W^2-s_W^2\right)\lambda^{\varphi_i^0}_{H^+H^-}\lambda^{\varphi_i^0}_{W^\pm H^\mp} \left(C_{22}-C_{23}\right)$&$\left[p_W,-p_{H^\pm},m_{H^\pm},m_{\varphi_i^0},m_{H^\pm}\right]$ \\ \hline
%%%%%%%%%%%%%%%%%%%%%%%%%%%%%%%%%%%%%%%%%%%%%%%%%%%%%%%%%%%%%%%%%%%%%%%%%% 
 $F^{\ref{fig1}[3]}$&$-\frac{g^4v}{4c_W}\lambda^{\varphi_i^0}_{AZ}\lambda^{\varphi_i^0}_{W^\pm W^\mp}C_{24}$ & $\left[p_W,-p_{H^\pm},m_{\varphi_i^0},m_{W^\pm},m_{A}\right]$ \\
$G^{\ref{fig1}[3]}$&$-\frac{g^4v}{4c_W}\lambda^{\varphi_i^0}_{AZ}\lambda^{\varphi_i^0}_{W^\pm W^\mp}\left(2(C_{12}-C_{11})+C_{22}-C_{23}\right)$ & $\left[p_W,-p_{H^\pm},m_{\varphi_i^0},m_{W^\pm},m_{A}\right]$\\\hline
%%%%%%%%%%%%%%%%%%%%%%%%%%%%%%%%%%%%%%%%%%%%%%%%%%%%%%%%%%%%%%%%%%%%%%%%%
 $F^{\ref{fig1}[4]}$&$\frac{g^4v}{4c_W^3}\left(c_W^2-s_W^2\right)\lambda^{\varphi_i^0}_{ZZ}\lambda^{\varphi_i^0}_{H^\pm W^\mp}C_{24}$ & $\left[p_W,-p_{H^\pm},m_{\varphi_i^0},m_{H^\pm},m_{Z}\right]$ \\
$G^{\ref{fig1}[4]}$&$\frac{g^4v}{4c_W^3}\left(c_W^2-s_W^2\right)\lambda^{\varphi_i^0}_{ZZ}\lambda^{\varphi_i^0}_{H^\pm W^\mp}\left(-2C_{12}+C_{22}-C_{23}\right)$ & $\left[p_W,-p_{H^\pm},m_{\varphi_i^0},m_{H^\pm},m_{Z}\right]$\\\hline
%%%%%%%%%%%%%%%%%%%%%%%%%%%%%%%%%%%%%%%%%%%%%%%%%%%%%%%%%%%%%%%%%%%%%%%%%
$F^{\ref{fig1}[5]}$&$-\frac{g^4c_Wv}{4}\lambda^{\varphi_i^0}_{W^\pm W^\mp}\lambda^{\varphi_i^0}_{H^\pm W^\mp}\big[(m_W^2-m_{H^\pm}^2)C_0
-\tilde{C}_0$& $\left[p_W,-p_{H^\pm},m_{W^\pm},m_{\varphi_i^0},m_{W^\pm}\right]$ \\
&$-(m_{H^\pm}^2+m_W^2-m_Z^2)C_{11}+2m_{H^\pm}^2C_{12}+C_{24}\big]$&  \\ 
$G^{\ref{fig1}[5]}$&$-\frac{g^4c_Wv}{4}\lambda^{\varphi_i^0}_{W^\pm W^\mp}\lambda^{\varphi_i^0}_{H^\pm W^\mp}\big[4(C_0+C_{11})-2C_{12}+C_{22}-
C_{23}\big]$ & $\left[p_W,-p_{H^\pm},m_{W^\pm},m_{\varphi_i^0},m_{W^\pm}\right]$\\\hline
%%%%%%%%%%%%%%%%%%%%%%%%%%%%%%%%%%%%%%%%%%%%%%%%%%%%%%%%%%%%%%%%%%%%%%5
$F^{\ref{fig1}[6]}$&$\frac{g^4v}{4c_W}\lambda^{\varphi_i^0}_{ZZ}\lambda^{\varphi_i^0}_{H^\pm W^\mp}\big[(m_Z^2-m_{H^\pm}^2)C_0
-\tilde{C}_0$& $\left[p_W,-p_{H^\pm},m_{Z},m_{W^\pm},m_{\varphi_i^0}\right]$ \\
&$+(m_{H^\pm}^2+m_W^2-m_Z^2)C_{11}-2m_{H^\pm}^2C_{12}+C_{24}\big]$&  \\ 
$G^{\ref{fig1}[6]}$&$\frac{g^4v}{4c_W}\lambda^{\varphi_i^0}_{ZZ}\lambda^{\varphi_i^0}_{H^\pm W^\mp}\big[4C_0+2(C_{11}+C_{12})+C_{22}-
C_{23}\big]$ & $\left[p_W,-p_{H^\pm},m_Z,m_{W^\pm},m_{\varphi_i^0}\right]$\\\hline
%%%%%%%%%%%%%%%%%%%%%%%%%%%%%%%%%%%%%%%%%%%%%%%%%%%%%%%%%%%%%%%%%%%%%%%%
$F^{\ref{fig1}[7]}$&$\frac{g^2v}{2c_W}(\lambda_4-2\lambda_5)\lambda^{\varphi_i^0}_{AZ}\lambda^{\varphi_i^0}_{G^\pm W^\mp}C_{24}$& $\left[p_W,-p_{H^\pm},m_{\varphi_i^0},m_{W^\pm},m_{A}\right]$ \\
$G^{\ref{fig1}[7]}$&$\frac{g^2v}{2c_W}(\lambda_4-2\lambda_5)\lambda^{\varphi_i^0}_{AZ}\lambda^{\varphi_i^0}_{G^\pm W^\mp}(C_{22}-C_{23})$& $\left[p_W,-p_{H^\pm},m_{\varphi_i^0},m_{W^\pm},m_{A}\right]$ \\ \hline
%%%%%%%%%%%%%%%%%%%%%%%%%%%%%%%%%%%%%%%%%%%%%%%%%%%%%%%%%%%%%%%%%%%%%%5
$F^{\ref{fig1}[8]}$&$-\frac{g^2v}{c_W}\lambda^{\varphi_i^0}_{G^0Z}\lambda^{\varphi_i^0}_{G^\pm H^\mp}C_{24}$&$\left[p_W,-p_{H^\pm},m_{Z},m_{W^\pm},m_{\varphi_i^0}\right]$ \\
$G^{\ref{fig1}[8]}$&$-\frac{g^2v}{c_W}\lambda^{\varphi_i^0}_{G^0Z}\lambda^{\varphi_i^0}_{G^\pm H^\mp}(C_{22}-C_{23})$&$\left[p_W,-p_{H^\pm},m_{Z},m_{W^\pm},m_{\varphi_i^0}\right]$ \\ \hline
%%%%%%%%%%%%%%%%%%%%%%%%%%%%%%%%%%%%%%%%%%%%%%%%%%%%%%%%%%%%%%%%%%%%%%
$F^{\ref{fig1}[9]}$&$\frac{g^2v}{c_W}(c_W^2-s_W^2)\lambda^{\varphi_i^0}_{G^\pm H^\mp}\lambda^{\varphi_i^0}_{G^\pm W^\mp}C_{24}$&$\left[p_W,-p_{H^\pm},m_{W^\pm},m_{\varphi_i^0},m_{W^\pm}\right]$ \\
$G^{\ref{fig1}[9]}$&$\frac{g^2v}{c_W}(c_W^2-s_W^2)\lambda^{\varphi_i^0}_{G^\pm H^\mp}\lambda^{\varphi_i^0}_{G^\pm W^\mp}(C_{22}-C_{23})$&$\left[p_W,-p_{H^\pm},m_{W^\pm},m_{\varphi_i^0},m_{W^\pm}\right]$ \\\hline 
%%%%%%%%%%%%%%%%%%%%%%%%%%%%%%%%%%%%%%%%%%%%%%%%%%%%%%%%%%%%%%%%%%%%%%%
$F^{\ref{fig1}[10]}$&$\frac{g^4v^3}{4c_W^3}s_W^2\lambda^{\varphi_i^0}_{G^\pm H^\mp}\lambda^{\varphi_i^0}_{ZZ}C_0$&$\left[p_W,-p_{H^\pm},m_{Z},m_{W^\pm},m_{\varphi_i^0}\right]$ \\\hline
%%%%%%%%%%%%%%%%%%%%%%%%%%%%%%%%%%%%%%%%%%%%%%%%%%%%%%%%%%%%%%%%%%%%%%%%
$F^{\ref{fig1}[11]}$&$\frac{g^4v^3}{4c_W}s_W^2\lambda^{\varphi_i^0}_{G^\pm H^\mp}\lambda^{\varphi_i^0}_{W^\pm W^\mp}C_0$&$\left[p_W,-p_{H^\pm},m_{W^\pm},m_{\varphi_i^0},m_{W^\pm}\right]$ \\\hline
%%%%%%%%%%%%%%%%%%%%%%%%%%%%%%%%%%%%%%%%%%%%%%%%%%%%%%%%%%%%%%%%%%%%%%%%%%%
$F^{\ref{fig1}[12]}$&$\frac{g^4v}{4c_W}s_W^2\lambda^{\varphi_i^0}_{G^\pm W\mp}\lambda^{\varphi_i^0}_{H^\pm W^\mp}C_{24}$&$\left[p_W,-p_{H^\pm},m_{W^\pm},m_{\varphi_i^0},m_{W^\pm}\right]$ \\
$G^{\ref{fig1}[12]}$&$\frac{g^4v}{4c_W}s_W^2\lambda^{\varphi_i^0}_{G^\pm W\mp}\lambda^{\varphi_i^0}_{H^\pm W^\mp}(-2C_{12}+C_{22}-C_{23})$&$\left[p_W,-p_{H^\pm},m_{W^\pm},m_{\varphi_i^0},m_{W^\pm}\right]$ \\ \hline
\end{tabular}
\caption{ \sl The $F$ and $G$ terms from the boson triangle loop diagrams Fig.~\ref{fig1} contributing to the decay $H^+\rightarrow W^+Z$, with $\varphi_1^0 = h$ and $\varphi_2^0 = H$.}
\label{tab:coupling1} 
\end{table}
%%%%%%%%%%%%%%%%%%%%%%%%%%%%%%%%%%%%%%%%%%%%%%%%%%%%%%%%%%%%%%%%%%%%%%
%%%%%%%%%%%%%%%%%%%%%%%%%%%%%%%%%%%%%%%%%%%%%%%%%%%%%%%%%%%%%%%%%%%%%%%%%%%%%%%%%%%%%%%%%%%%%
\begin{table}[!h]
\centering
\begin{tabular}{|c|c|c|}
\hline 
Fig.~\ref{fig2}[n] &amplitude & argument \\ \hline
$F^{\ref{fig2}[1]}_{H^\pm}$&$\frac{g^2v}{2c_W}s_W^2 \lambda^{\varphi_i^0}_{H^\pm H\mp}\lambda^{\varphi_i^0}_{H^\pm W^\mp Z} B_0$ &$\left[p_{H^\pm},m_{\varphi_i^0},m_{H^\pm}\right]$ \\ \hline
%%%%%%%%%%%%%%%%%%%%%%%%%%%%%%%%%%%%%%%%%%%%%%%%%%%%%%%%%%%%%%%%%%%%%
$F^{\ref{fig2}[1]}_{G^\pm}$&$\frac{g^2v}{2c_W}s_W^2 \lambda^{\varphi_i^0}_{G^\pm H\mp}\lambda^{\varphi_i^0}_{G^\pm W^\mp Z} B_0$ &$\left[p_{H^\pm},m_{\varphi_i^0},m_{W^\pm}\right]$ \\ \hline
%%%%%%%%%%%%%%%%%%%%%%%%%%%%%%%%%%%%%%%%%%%%%%%%%%%%%%%%%%%%%%%%%%%%%%%%
$F^{\ref{fig2}[2]}$&$\frac{g^4v}{4c_W}s_W^2 \lambda^{\varphi_i^0}_{W^\pm W\mp}\lambda^{\varphi_i^0}_{H^\pm W^\mp Z} B_0$ &$\left[-p_{W},m_{W^\pm},m_{\varphi_i^0}\right]$ \\ \hline
%%%%%%%%%%%%%%%%%%%%%%%%%%%%%%%%%%%%%%%%%%%%%%%%%%%%%%%%%%%%%%%%%%%%%%%
$F^{\ref{fig2}[3]}$&$\frac{g^4v}{4c_W^3}s_W^2 \lambda^{\varphi_i^0}_{ZZ}\lambda^{\varphi_i^0}_{H^\pm W^\mp Z} B_0$ &$\left[-p_{Z},m_{\varphi_i^0},m_{Z}\right]$ \\ \hline
%%%%%%%%%%%%%%%%%%%%%%%%%%%%%%%%%%%%%%%%%%%%%%%%%%%%%%%%%%%%%%%%%%%%%%%%%%%
$F^{\ref{fig2}[4]}$&$-\frac{g^4c_W v}{4}\frac{m_W^2-m_Z^2}{m_{H^{\pm}}^2-m_W^2} \lambda^{\varphi_i^0}_{W^\pm W^\mp}\lambda^{\varphi_i^0}_{H^\pm W^\mp} (B_0-B_1)$ &$\left[p_{H^\pm},m_{\varphi_i^0},m_{W^\pm}\right]$ \\ \hline
%%%%%%%%%%%%%%%%%%%%%%%%%%%%%%%%%%%%%%%%%%%%%%%%%%%%%%%%%%%%%%%%%%%%%%%%%%
$F^{\ref{fig2}[5]}_{H^\pm}$&$-\frac{g^2c_W v}{2}\frac{m_W^2-m_Z^2}{m_{H^{\pm}}^2-m_W^2} \lambda^{\varphi_i^0}_{H^\pm H^\mp}\lambda^{\varphi_i^0}_{H^\pm W^\mp} (B_0+2B_1)$ &$\left[p_{H^\pm},m_{\varphi_i^0},m_{H^\pm}\right]$ \\ \hline
%%%%%%%%%%%%%%%%%%%%%%%%%%%%%%%%%%%%%%%%%%%%%%%%%%%%%%%%%%%%%%%%%%%%%%%%%
$F^{\ref{fig2}[5]}_{G^\pm}$&$-\frac{g^2c_W v}{2}\frac{m_W^2-m_Z^2}{m_{H^{\pm}}^2-m_W^2} \lambda^{\varphi_i^0}_{G^\pm H^\mp}\lambda^{\varphi_i^0}_{G^\pm W^\mp} (B_0+2B_1)$ &$\left[p_{H^\pm},m_{\varphi_i^0},m_{W^\pm}\right]$ \\ \hline
%%%%%%%%%%%%%%%%%%%%%%%%%%%%%%%%%%%%%%%%%%%%%%%%%%%%%%%%%%%%%%%%%%%%%%%%
$F^{\ref{fig2}[6]}$&$-\frac{g^4v}{8c_W}\frac{s_W^2}{m_{H^{\pm}}^2-m_W^2} \lambda^{\varphi_i^0}_{G^\pm W^\mp}\lambda^{\varphi_i^0}_{H^\pm W^\mp} (m_{H^\pm}^2 (B_0-2B_1)+\tilde{B}_0)$ &$\left[p_{H^\pm},m_{\varphi_i^0},m_{W^\pm}\right]$ \\ \hline
%%%%%%%%%%%%%%%%%%%%%%%%%%%%%%%%%%%%%%%%%%%%%%%%%%%%%%%%%%%%%%%%%%%%%%%%
$F^{\ref{fig2}[7]}_{H^\pm}$&$\frac{g^2v^3}{2c_W}\frac{s_W^2}{m_{H^{\pm}}^2-m_W^2} \lambda^{\varphi_i^0}_{H^\pm H^\mp}\lambda^{\varphi_i^0}_{G^\pm H^\mp} B_0$&$\left[p_{H^\pm},m_{\varphi_i^0},m_{H^\pm}\right]$ \\ \hline
%%%%%%%%%%%%%%%%%%%%%%%%%%%%%%%%%%%%%%%%%%%%%%%%%%%%%%%%%%%%%%%%%%%%%%%%%
$F^{\ref{fig2}[7]}_{G^\pm}$&$\frac{g^2v^3}{2c_W}\frac{s_W^2}{m_{H^{\pm}}^2-m_W^2} \lambda^{\varphi_i^0}_{G^\pm H^\mp}\lambda^{\varphi_i^0}_{G^\pm G^\mp} B_0$&$\left[p_{H^\pm},m_{\varphi_i^0},m_{W^\pm}\right]$ \\ \hline
%%%%%%%%%%%%%%%%%%%%%%%%%%%%%%%%%%%%%%%%%%%%%%%%%%%%%%%%%%%%%%%%%%%%%%%%%
$F^{\ref{fig2}[8]}$& $-\frac{v}{m_{\varphi_i^0}^2} \frac{g^2s_W^2}{4 c_W} \lambda^{\varphi_i^0}_{H^\pm W^\mp Z} \bigg[4g^2 \lambda^{\varphi_i^0}_{W^\pm W^\mp}A_0[m_{W^\pm}]+\frac{2g^2}{c_W^2}\lambda^{\varphi_i^0}_{ZZ}A_0[m_{Z}]$ &\\ 
& $+\lambda^{\varphi_i^0}_{\mathcal{P}\mathcal{P}} A_0[m_{\mathcal{P}}] + 2\left( \lambda^{\varphi_i^0}_{G^\pm G^\pm} A_0[m_{W^\pm}]+\lambda^{\varphi_i^0}_{H^\pm H^\pm} A_0[m_{H^\pm}]\right)\bigg]$ & \\ \hline
%%%%%%%%%%%%%%%%%%%%%%%%%%%%%%%%%%%%%%%%%%%%%%%%%%%%%%%%%%%%%%%%%%%%%%%%%
$F^{\ref{fig2}[9]}$ & $\frac{g^2 v}{c_W}\frac{s_W^2}{m_{H^\pm}^2-m_{W^\pm}^2}
\bigg[\lambda_7 \left(A_0[m_{H^\pm}]+\frac{A_0[m_A]}{4}\right)$ & \\ 
&
$+\frac{1}{4}\lambda^{\varphi_i^0\varphi_i^0}_{G^\pm H^\pm} A_0[m_{\varphi_i^0}] +\lambda_6 \left(A_0[m_{W^\pm}]+\frac{A_0[m_Z]}{4}\right)\bigg]$ & \\ \hline
%%%%%%%%%%%%%%%%%%%%%%%%%%%%%%%%%%%%%%%%%%%%%%%%%%%%%%%%%%%%%%%%%%%%%%%%%%%%%%%%%%%%
$F^{\ref{fig2}[10]}$& $-\frac{g^2vc_W}{2m_{\varphi_i^0}^2}\frac{m_{W^\pm}^2-m_Z^2}{m_{H^\pm}^2-m_{W^\pm}^2}R_{i2}\bigg[-2 g^2 \lambda^{\varphi_i^0}_{W^\pm W^\mp} A_0[m_{W^\pm}] - \frac{g^2}{c_W^2}\lambda^{\varphi_i^0}_{ZZ} A_0[m_{Z}]$ & \\ 
& $-\frac{1}{2}\lambda^{\varphi_i^0}_{\mathcal{P}\mathcal{P}}
A_0[m_{\mathcal{P}}]-\left(\lambda^{\varphi_i^0}_{G^\pm G^\pm} A_0[m_{W^\pm}+\lambda^{\varphi_i^0}_{H^\pm H^\pm} A_0[m_{H^\pm}
\right)]\bigg] $ & \\ \hline
%%%%%%%%%%%%%%%%%%%%%%%%%%%%%%%%%%%%%%%%%%%%%%%%%%%%%%%%%%%%%%%%%%%%%%%%%%%%%%%%%%%%%%
$F^{\ref{fig2}[11]}$ & $\frac{g^2 v^3}{2c_W}\frac{s_W^2}{m_{H^\pm}^2-m_{W^\pm}}\frac{1}{m_{\varphi_i^0}^2}\bigg[ 2g^2 \lambda^{\varphi_i^0}_{W^\pm W^\mp}A_0[m_{W^\pm}]+\frac{g^2}{c_W^2}\lambda^{\varphi_i^0}_{ZZ} A_0[m_{Z}]$ & \\ 
& $ +\frac{1}{2}\lambda^{\varphi_i^0}_{\mathcal{P}\mathcal{P}}
A_0[m_{\mathcal{P}}]+\left(\lambda^{\varphi_i^0}_{G^\pm G^\pm} A_0[m_{W^\pm}+\lambda^{\varphi_i^0}_{H^\pm H^\pm} A_0[m_{H^\pm}
\right)]\bigg]$ & \\ \hline
%%%%%%%%%%%%%%%%%%%%%%%%%%%%%%%%%%%%%%%%%%%%%%%%%%%%%%%%%%%%%%%%%%%%%%%%%%%%%%%%%%%%%%%%
\end{tabular}
\caption{ \sl The $F$ term from the boson loop diagrams Fig.~\ref{fig2} contributing to the decay $H^+\rightarrow W^+Z$, where~$\mathcal{P}=h,~H,~A,~G^0$  and $\varphi_1^0 = h$, $\varphi_2^0 = H$ }
\label{tab:coupling2} 
\end{table}
%%%%%%%%%%%%%%%%%%%%%%%%%%%%%%%%%%%%%%%%%%%%%%%%%%%%%%%%%%%%%%%%%%%%%%%%%%%%%%%%%%%%%%%%%%%%%%%%%%%%%%%%%%%%%%%%%%%%%%%%%%
\begin{table}[!h]
\centering
\begin{tabular}{|c|c|c|}
\hline 
Fig.~\ref{fig3}[n] &amplitude & argument \\ \hline
$F^{\ref{fig3}[1]}$&$- \frac{2N_C g^2 c_W}{v}\frac{m_Z^2-m_{W^\pm}^2}{m_{H^\pm}^2-m_{W^\pm}^2}|V_{ud}|^2\left\lbrace m_d^2\varsigma_d (B_1+B_0)-m_u^2\varsigma_u B_1\right\rbrace$ &$\left[p_{H^\pm},m_{d},m_{u}\right]$ \\ \hline
%%%%%%%%%%%%%%%%%%%%%%%%%%%%%%%%%%%%%%%%%%%%%%%%%%%%%%%%%%
&$\frac{2N_Cg^2}{v c_W}|V_{ud}|^2 \left\lbrace m_d^2(g^d_V-g^d_A)\left(\frac{\varsigma_d}{2}(m_{H^\pm}^2-m_{W^\pm}^2-m_{Z}^2)+m_u^2 \varsigma_u\right) C_0\right.$&\\
&$\left. +\left(2g^d_A m_d^2 \varsigma_d-(g^d_A+g^d_V)m_u^2 \varsigma_u\right)\tilde{C}_0\right.$&\\
&$\left.
-\frac{1}{2}\left(g^d_A(m_d^2 \varsigma_d-m_u^2 \varsigma_u)-g^d_V(m_d^2 \varsigma_d+m_u^2 \varsigma_u)\right)\right.$&\\
$F^{\ref{fig3}[2]}$&$\left.\left((m_{H^\pm}^2-m_{W^\pm}^2-m_{Z}^2)C_{11}-(m_{H^\pm}^2-m_{W^\pm}^2+m_{Z}^2)C_{12}\right)\right.$&$\left[p_W,-p_{H^\pm},m_d,m_u,m_d\right]$\\
&$\left.
+2g^d_A m_d^2 \varsigma_d \left(m_{W^\pm}^2 C_{11}-\frac{1}{2}(m_{H^\pm}^2+m_{W^\pm}^2-m_{Z}^2)C_{12}\right)\right.$&\\
&$\left.
-2(g^d_A+g^d_V)(m_d^2 \varsigma_d-m_u^2\varsigma_u)C_{24}\right\rbrace$& \\ \hline
$G^{\ref{fig3}[2]}$&$\frac{2N_Cg^2}{v c_W}|V_{ud}|^2\left\lbrace m_d^2 \varsigma_d (g^d_A-g^d_V)(C_0+C_{11})-m_u^2\varsigma_u(g^d_A+g^d_V)C_{11}\right. $&$\left[p_W,-p_{H^\pm},m_d,m_u,m_d\right]$\\
&$\left.+(g^d_A+g^d_V)\left((m_d^2 \varsigma_d+m_u^2\varsigma_u)C_{12}-2(m_d^2 \varsigma_d-m_u^2\varsigma_u)(C_{22}-C_{23})\right)\right\rbrace$ &\\ \hline
$H^{\ref{fig3}[2]}$&$\frac{2N_Cg^2}{v c_W}|V_{ud}|^2\left\lbrace m_d^2 \varsigma_d (g^d_A-g^d_V)(C_0+C_{11})+m_u^2\varsigma_u(g^d_A+g^d_V)C_{11}\right.$&$\left[p_W,-p_{H^\pm},m_d,m_u,m_d\right]$\\
&$\left.+(g^d_A+g^d_V)(m_d^2 \varsigma_d-m_u^2\varsigma_u)C_{12}\right\rbrace$ &\\ \hline
%%%%%%%%%%%%%%%%%%%%%%%%%%%%%%%%%%%%%%%%%%%%%%%%%%%%%%%%%%%%%
%%%%%%%%%%%%%%%%%%%%%%%%%%%%%%%%%%%%%%%%%%%%%%%%%%%%%%%%%%%%%%
%%%%%%%%%%%%%%%%%%%%%%%%%%%%%%%%%%%%%%%%%%%%%%%%%%%%%%%%%%%%%%
&$\frac{2N_Cg^2}{v c_W}|V_{ud}|^2\left\lbrace m_u^2 (g^u_A-g^u_V)\left(m_d^2 \varsigma_d+\frac{\varsigma_u}{2}(m_{H^\pm}^2-m_{W^\pm}^2-m_Z^2)\right)C_0\right.$& \\
&$\left.+\left(m_d^2\varsigma_d(g^u_A+g^u_V)-2m_u^2\varsigma_u g^u_A\right)\tilde{C}_0\right.$& \\
&$\left.-\frac{1}{2}(m_d^2 \varsigma_d(g^u_A+g^u_V)-m_u^2 \varsigma_u(g^u_A-g^u_V))\right.$& \\
$F^{\ref{fig3}[3]}$&$\left.\left((m_{H^\pm}^2-m_{W^\pm}^2-m_Z^2)C_{11}-(m_{H^\pm}^2-m_{W^\pm}^2+m_Z^2)C_{12}\right)\right.$&$\left[p_W,-p_{H^\pm},m_u,m_d,m_u\right]$ \\
&$\left.-2g^u_A m_u^2 \varsigma_u \left(m_{W^\pm}^2C_{11}-\frac{1}{2}(m_{H^\pm}^2+m_{W^\pm}^2-m_Z^2))C_{12}\right)\right.$& \\
&$\left.-2(g^u_A+g^u_V)(m_d^2 \varsigma_d-m_u^2 \varsigma_u)C_{24}\right\rbrace$ &\\ \hline
%%%%%%%%%%%%%%%%%%%%%%%%%%%%%%%%%%%%%%%%%%%%%%%%%%%%%%%%%%%%%
$G^{\ref{fig3}[3]}$&$\frac{2 N_C g^2}{v c_W}|V_{ud}|^2\left\lbrace m_u^2 \varsigma_u (g^u_V-g^u_A)(C_0+C_{11})+m_d^2\varsigma_d(g^u_A+g^u_V)C_{11}\right.$&\\
&$\left.-(g^u_A+g^u_V)\left((m_d^2 \varsigma_d+m_u^2 \varsigma_u)C_{12}+2(m_d^2\varsigma_d-m_u^2 \varsigma_u)(C_{22}-C_{23})\right)\right\rbrace$ &$\left[p_W,-p_{H^\pm},m_u,m_d,m_u\right]$ \\ \hline
%%%%%%%%%%%%%%%%%%%%%%%%%%%%%%%%%%%%%%%%%%%%%%%%%%%%%%%%%%%%%
$H^{\ref{fig3}[3]}$&$\frac{2 N_C g^2}{v c_W}|V_{ud}|^2\left\lbrace m_u^2 \varsigma_u (g^u_A-g^u_V)(C_0+C_{11})+m_d^2\varsigma_d(g^u_A+g^u_V)C_{11}\right.$&\\
&$\left.-(g^u_A+g^u_V)\left((m_d^2 \varsigma_d-m_u^2 \varsigma_u)C_{12}\right)\right\rbrace$ &$\left[p_W,-p_{H^\pm},m_u,m_d,m_u\right]$ \\ \hline
%%%%%%%%%%%%%%%%%%%%%%%%%%%%%%%%%%%%%%%%%%%%%%%%%%%%%%%%%%%%
%%%%%%%%%%%%%%%%%%%%%%%%%%%%%%%%%%%%%%%%%%%%%%%%%%%%%%%%%%%%
%%%%%%%%%%%%%%%%%%%%%%%%%%%%%%%%%%%%%%%%%%%%%%%%%%%%%%%%%%%%
$F^{\ref{fig3}[4]}$    &$- \frac{2N_C g^2}{v c_W}\frac{ s_W^2}{m_{H^\pm}^2-m_{W^\pm}^2}|V_{ud}|^2\left\lbrace \left(m_d^2\varsigma_d+m_u^2\varsigma_u\right) (m_{H^\pm}^2 B_1+\tilde{B}_0)\right.$& $\left[p_{H^\pm},m_{d},m_{u}\right]$ \\
&$\left.-m_u^2 m_d^2(\varsigma_u+\varsigma_d) B_0\right\rbrace$ & \\ \hline
%%%%%%%%%%%%%%%%%%%%%%%%%%%%%%%%%%%%%%%%%%%%%%%%%%%%%%%%%%%%%%%%%%
%%%%%%%%%%%%%%%%%%%%%%%%%%%%%%%%%%%%%%%%%%%%%%%%%%%%%%%%%%%%%%%%%%%%
$F^{\ref{fig3}[5]}$ &  $\frac{6}{m_{\varphi_i^0}^2} \frac{g^2 s_W^2}{v c_W} \lambda^{\varphi_i^0}_{H^\pm W^\mp Z} \sum\limits_{f = u,d} ( m_f^2  \lambda^{\varphi_i^0}_{f\bar{f}}   A_0) $ 
   &$m_{u,d}$  \\ \hline
 %%%%%%%%%%%%%%%%%%%%%%%%%%%%%%%%%%%%%%%%%%%%%%%%%%%%%%%%%%%%%%%%%%
%%%%%%%%%%%%%%%%%%%%%%%%%%%%%%%%%%%%%%%%%%%%%%%%%%%%%%%%%%%%%%%%%%%%
$F^{\ref{fig3}[6]}$ &  $\frac{6}{m_{\varphi_i^0}^2} \frac{g^2 c_W}{v}  \frac{m_Z^2-m_W^2}{m_{H^\pm}^2-m_W^2} \lambda^{\varphi_i^0}_{H^\pm W^\mp }   \sum\limits_{f = u,d} ( m_f^2  \lambda^{\varphi_i^0}_{f\bar{f}}   A_0) $ 
   &$m_{u,d}$  \\ \hline
 %%%%%%%%%%%%%%%%%%%%%%%%%%%%%%%%%%%%%%%%%%%%%%%%%%%%%%%%%%%%%%%%%%
%%%%%%%%%%%%%%%%%%%%%%%%%%%%%%%%%%%%%%%%%%%%%%%%%%%%%%%%%%%%%%%%%%%%
$F^{\ref{fig3}[7]}$ &  $\frac{6}{m_{\varphi_i^0}^2} \frac{g^2  v s_W^2}{c_W} \frac{1 }{m_{H^\pm}^2-m_W^2}\lambda^{\varphi_i^0}_{H^\pm G^\mp }\sum\limits_{f = u,d} ( m_f^2  \lambda^{\varphi_i^0}_{f\bar{f}}   A_0) $ 
   &$m_{u,d}$  \\ \hline  
\end{tabular}
\caption{ \sl The $F$, $G$ and $H$ term from the fermion loop diagrams Fig.~\ref{fig3} contributing to the decay $H^+\rightarrow W^+Z$,  with $\varphi_1^0 = h$ and $\varphi_2^0 = H$. Here $u$ represents all the up-type quarks and $d$ represents the down-type quarks and the fermions.}
\label{tab:coupling3} 
\end{table}

\section*{Acknowledgments}
We are extremely grateful to Antonio Pich and Saurabh D. Rindani for  reading the manuscript thoroughly,  very useful discussions, comments, and suggestions on the manuscript.  D.D. is supported by the DST, Government of India under INSPIRE Faculty Award (letter number DST/INSPIRE/04/2016/002620). M.P. acknowledges support of the Slovenian Research Agency through research core funding No. P1-0035.

\appendix

\section{Amplitudes for $H^\pm \rightarrow  W^\pm Z$ for Feynman diagrams given in diagrams Figs.~\ref{fig1}, \ref{fig2} and \ref{fig3}  }\label{sec:FGHterms}
We present in this section the $F$, $G$ and $H$ form factors used to parametrize the one-loop decay amplitude defined in Eq.~(\ref{eq:mat_element}).  We would like to point out that the fermion-loop contribution to the $H^+W^-Z$ vertex was earlier calculated in the unitary gauge~\cite{Kanemura:1997ej}. The fermion-loop diagrams by themselves form a gauge invariant subset whereas the boson-loop diagrams form another subset so that they can be independently calculated in an arbitrary gauge.  We perform the calculation for the contributions from the boson and the fermion-loop diagrams in the 't Hooft-Feynman gauge and have explicitly checked that our results are finite and independent of the gauge parameter. The $F$, $G$ and  $H$  terms are listed in Tables~\ref{tab:coupling1},~\ref{tab:coupling2} and~\ref{tab:coupling3}. Here we show the contributions to $F, G$ and $H$ separately for each diagrams. The notation is as follows: $F^{\ref{fig1}[1]}$ indicates the contribution of the first diagram in Fig.~\ref{fig1} and so on. The same definition holds for the others. We list below the various couplings used for the computation, 
\begin{eqnarray}
\lambda^{\varphi_i^0}_{H^\pm H^\mp}&=&\lambda_3 R_{i1} +\lambda_7 R_{i2} ,\nonumber \\
\lambda^{\varphi_i^0}_{AZ}&=& \lambda^{\varphi_i^0}_{H^\pm W^\mp}=\lambda^{\varphi_i^0}_{H^\pm W^\mp Z}=R_{i2} ,\nonumber \\
\lambda^{\varphi_i^0}_{W^\pm W^\mp}&=&\lambda^{\varphi_i^0}_{ZZ}=\lambda^{\varphi_i^0}_{G^\pm W^\mp}=\lambda^{\varphi_i^0}_{G^0Z}=\lambda^{\varphi_i^0}_{G^\pm W^\mp Z}= R_{i1} ,\nonumber \\
\lambda^{\varphi_i^0}_{G^\pm H^\mp}&=&\lambda_6 R_{i1}+\left(\frac{\lambda_4}{2}+\lambda_5\right)R_{i2},\nonumber \\
\lambda^{\varphi_i^0}_{G^\pm G^\mp}&=& \lambda^{\varphi_i^0}_{G^0 G^0} = 2 \lambda_1 R_{i1}+\lambda_6 R_{i2},\nonumber \\
\lambda^{\varphi_i^0}_{A A}&=& (\lambda_3+\lambda_4-2 \lambda_5) R_{i1}+\lambda_7 R_{i2},\nonumber \\
\lambda^{\varphi_i^0}_{\varphi_i^0\varphi_i^0}&=&2\lambda_1 R_{i1}^3+R_{i2}^2R_{i1}(\lambda_3+\lambda_4+2\lambda_5)+3\lambda_6 R_{i2}R_{i1}^2+\lambda_7 R_{i2}^3 \nonumber \\
\lambda^{\varphi_i^0}_{\varphi_j^0\varphi_j^0}&=&6 \lambda_1(-  R_{i1}^3+ R_{i1}+3 R_{i1}R_{i2}^2)+(\lambda_3 + \lambda_4) (3 R_{i1}^3+R_{i1}-9R_{i2}^2R_{i1}) \nonumber \\
&&+2 \lambda_5( 3 R_{i1}^3+ R_{i1}-9 R_{i1}R_{i2}^2)+3 \lambda_6 (3R_{i2}^3+ R_{i2}-9 R_{i1}^2R_{i2}) \nonumber \\
&&+3\lambda_7(- R_{i2}^3+ R_{i2}+3 R_{i1}^2R_{i2}) ~~\mathrm{for}~i\neq j,\nonumber \\
\lambda^{\varphi_i^0}_{f\bar{f}}&=& R_{i1} + \varsigma_f R_{i2}
\end{eqnarray}
where $R_{ij}$ denotes the $ij$ element of the orthogonal $R$ matrix, which determines the neutral CP-even Higgs boson mass eigenstates defined in Eq.~\ref{eq:rot_matrix}.
\section{Loop integrals}
\label{app:III}

We have used the dimensional regularization scheme for our calculation, with the approach similar to one given in the appendix of Ref.~\cite{Axelrod:1982yc}. The integration measure in $D$  dimension is given by 
\be
d^{D} \tilde k\; =\; \mu^{ 3(4-D)/2}\,   \dfrac{d^{D} k}{   (2 \pi)^{D} }\,,
\ee
where $g \mu^{(4-D)/2}$ is the $\mathrm{SU(2)}_L$ gauge coupling constant in $D$ dimensions. The scalar loop functions appearing are given by~\cite{'tHooft:1978xw,Passarino:1978jh}
\begin{align}
A_0(m_1) &\;=\; \int d^{D} \tilde k \;\dfrac{1}{  k^2 -m_1^2  }\,, \nonumber \\[0.2cm]
B_0(l, m_1, m_2) &\;=\; \int d^{D} \tilde k \;\dfrac{1}{    (k^2 -m_1^2)   [    (k +l)^2 - m_2^2 ]  } \,, \nonumber \\[0.2cm]
C_0(l, s, m_1, m_2, m_3 ) &\;=\;   \int d^{D} \tilde k \;\dfrac{1}{    (k^2 - m_1^2)    [   (k+l)^2  -m_2^2 ]   [   (k+l +s)^2 -m_3^2     ]}  \,.
\end{align}
%
%The vector and tensor integrals are reduced to the scalar loop integrals via the Passarino--Veltman method~\cite{Passarino:1978jh}. 
We use the definitions $s_1= l^2 + m_1^2 -m_2^2$ and $s_2= s^2 + 2 l \cdot s + m_2^2 - m_3^2 $ following Ref.~\cite{Axelrod:1982yc}. The loop functions are given by 
\begin{align}
 \widetilde B_0 &\;=\;  A_0(m_2)  + m_1^2 B_0 \,, \quad 
B_1 \;=\; \frac{1}{2 l^2}   \bigl[ A_0(m_1) - A_0(m_2)  - s_1 B_0  \bigr] \,, \nonumber \\[0.2cm]
\widetilde C_0 &\;=\;   B_0(s,m_2,m_3)  + m_1^2 C_0 \,,
\end{align}
\begin{align}
C_{22} \;=\; & \dfrac{1}{  2 \bigl[  l^2 s^2 - (l s)^2 \bigr] }\; \biggl\{  - l s \Bigl[ B_1(l+s,m_1,m_3)    -B_1(s,m_2,m_3)    - s_1 C_{12}   \Bigr] \nonumber \\
&+ l^2 \Bigl[  -B_1(l+s,m_1,m_3) -s_2 C_{12}  -2 C_{24}      \Bigr]
\biggr\}    \,, \nonumber \\[0.3cm]
C_{24} \;=\; & \frac{1}{   2 ( D - 2) }   \Bigl[   1+  B_0(s,m_2,m_3)   + 2 m_1^2 C_0   + s_1 C_{11}  + s_2 C_{12}   \Bigr]   \,.
\end{align}
\begin{align}\label{eq:CsII}
\left(\ba C_{11}\\  C_{12}\ea\right)\; &= \; \frac{1}{2} \mathcal{X}
\left[\ba B_0(l +s,m_1, m_3) -B_0(s,m_2,m_3)  - s_1 C_0     \\[0.2cm]  B_0(l,m_1, m_2) -B_0(l+s,m_1,m_3)  - s_2 C_0 \ea\right] \,, \nonumber \\[0.3cm]
\left(\ba C_{21} \\  C_{23}\ea\right)\; &= \; \frac{1}{2} \mathcal{X} \left[\ba B_1(l +s,m_1, m_3) +B_0(s,m_2,m_3)  - s_1 C_{11} -2 C_{24}  \\[0.2cm]  B_1(l,m_1, m_2) -B_1(l+s,m_1,m_3)  - s_2 C_{11} \ea\right] \,,
\end{align}
with
\bel{eq:factor}
 \mathcal{X} = \; \dfrac{1}{  \bigl[ l^2 s^2 - (l s)^2 \bigr] }\,
\left[\bat   s^2   &   - l s  \\[0.2cm]
- l  s  &  l^2 \ea\right]\, .
\ee

\end{document}